\documentclass[]{emulateapj}
\usepackage{epstopdf}
\usepackage{multirow}
\usepackage{natbib}

\def\degs{\ifmmode ^{\circ}\else$^{\circ}$\fi}

\slugcomment{}

\shorttitle{Photometric redshift for {\it Chandra}-COSMOS}
\shortauthors{Salvato et al.}

\begin{document}


\title{Dissecting Photometric redshift for AGN using XMM- and {\it Chandra}- COSMOS samples}

\author{
M. Salvato\altaffilmark{1}, 
O. Ilbert\altaffilmark{2},
G. Hasinger\altaffilmark{1,3},
A. Rau\altaffilmark{4},
F. Civano\altaffilmark{5},
G. Zamorani\altaffilmark{6},
M. Brusa\altaffilmark{4},
M. Elvis\altaffilmark{5},
C. Vignali\altaffilmark{6,7},
H. Aussel\altaffilmark{8},
A. Comastri\altaffilmark{6},
F.  Fiore\altaffilmark{9},
E. Le Floc'h\altaffilmark{8},
V. Mainieri\altaffilmark{10},
S. Bardelli\altaffilmark{6},
M. Bolzonella\altaffilmark{6},
A. Bongiorno\altaffilmark{4}
P. Capak\altaffilmark{11},
K. Caputi\altaffilmark{12},
N. Cappelluti\altaffilmark{6},
C. M. Carollo\altaffilmark{13},
T. Contini\altaffilmark{14,15},
B. Garilli\altaffilmark{16},
A. Iovino\altaffilmark{16},
S. Fotopoulou\altaffilmark{1},
A. Fruscione\altaffilmark{5},
R. Gilli\altaffilmark{6},
C. Halliday\altaffilmark{17},
J-P. Kneib\altaffilmark{2},
Y. Kakazu\altaffilmark{18},
J.S. Kartaltepe\altaffilmark{19},
A. M. Koekemoer\altaffilmark{20},
K. Kovac\altaffilmark{21},
Y. Ideue\altaffilmark{22},
H. Ikeda\altaffilmark{22},
C.D. Impey\altaffilmark{23},
O. Le Fevre\altaffilmark{2},
F. Lamareille\altaffilmark{14,15},
G. Lanzuisi\altaffilmark{24},
J-F. Le Borgne\altaffilmark{14,15},
V. Le Brun\altaffilmark{14,15},
S. J. Lilly\altaffilmark{13},
C. Maier\altaffilmark{13},
S. Manohar\altaffilmark{18},
D. Masters\altaffilmark{18,24},
H. McCracken\altaffilmark{25},
H. Messias\altaffilmark{26},
M. Mignoli\altaffilmark{6},
B. Mobasher\altaffilmark{27},
T. Nagao\altaffilmark{22,28},
R. Pello\altaffilmark{14,15},
S. Puccetti\altaffilmark{29},
E. Perez Montero\altaffilmark{14,15,30}
A. Renzini\altaffilmark{31},
M. Sargent\altaffilmark{8},
D.B. Sanders\altaffilmark{3},
M. Scodeggio \altaffilmark{16},
N. Scoville\altaffilmark{18},
P. Shopbell\altaffilmark{18},
J. Silvermann\altaffilmark{32},
Y. Taniguchi\altaffilmark{33},
L. Tasca\altaffilmark{2},
L. Tresse\altaffilmark{2},
J.R. Trump\altaffilmark{34},
E. Zucca\altaffilmark{6}
   }
  \altaffiltext{$\star$}{Based on observations by the Chandra X-ray Observatory Center, which is operated by the Smithsonian Astrophysical Observatory for and on behalf of the National Aeronautics Space Administration under contract NAS8-03060.
 Also based  on observations with the NASA/ESA {\em Hubble Space Telescope}, obtained at the Space Telescope Science
Institute, which is operated by AURA Inc, under NASA contract NAS
5-26555. 
Also based on observations made with the Spitzer Space Telescope,
which is operated by the Jet Propulsion Laboratory, California Institute
of Technology, under NASA contract 1407.
 Also based on data collected at:
the Subaru Telescope, which is operated by the National Astronomical Observatory of Japan; 
the XMM-Newton, an ESA science mission with
instruments and contributions directly funded by ESA Member States and NASA; 
the European Southern Observatory under Large Program 175.A-0839,Chile;
the Kitt Peak National Observatory, Cerro Tololo Inter-American
Observatory and the National Optical Astronomy Observatory, which are
operated by the Association of Universities for Research in Astronomy, Inc.
(AURA) under cooperative agreement with the National Science Foundation;
 the Canada-France-Hawaii Telescope with MegaPrime/MegaCam operated as a
joint project by the CFHT Corporation, CEA/DAPNIA, the NRC and CADC of
Canada, the CNRS of France, TERAPIX and the Univ. of
Hawaii.}
\altaffiltext{1}{Max Planck Institut f\"ur Plasma Physik and Excellence Cluster,
      Boltzmannstrasse 2, 85748 Garching, Germany.}
      \email{mara.salvato@ipp.mpg.de}
\altaffiltext{2}{Laboratoire dÕAstrophysique de Marseille (UMR 6110) CNRS Universit\'e de Provence, 38 rue Fr\'ed\'eric Joliot-Curie, F-13388
Marseille Cedex 13, France }  
\altaffiltext{3}{Institute for Astronomy, University of  Hawaii, 2680 Woodlawn Drive,Honolulu, HI, 96822 USA.}
\altaffiltext{4}{Max Planck Institut f\"ur extraterrestrische Physik, Giessenbachstrasse 1, D--85748 Garching, Germany.}
\altaffiltext{5}{Harvard-Smithsonian Center for Astrophysics 60 Garden St., Cambridge, Massachusetts 02138 USA.}
\altaffiltext{6}{INAF--Osservatorio Astronomico di Bologna, via Ranzani 1, I--40127 Bologna, Italy.}
\altaffiltext{7}{Dipartimento di Astronomia Universita' di Bologna via Ranzani 1, I--40127 Bologna, Italy.}
\altaffiltext{8}{CEA/DSM-CNRS, Universit\'e Paris Diderot, IRFU/SAp, Orme des Merisiers, 91191, Gif-sur-Yvette, France.}
\altaffiltext{9}{INAF-Osservatorio Astronomico di Roma, via Frascati 33, Monteporzio Roma, Italy.}
\altaffiltext{10}{European Southern Observatory, Karl-Schwarzschild-str. 2, 85748 Garching, Germany.}
\altaffiltext{11}{Spitzer Science Center, California Institute of Technology, Pasadena, CA 91125.}
\altaffiltext{12}{SUPA, Institute for Astronomy, The University of Edinburgh, Royal Observatory, Edinburgh EH9 3HJ, UK}
\altaffiltext{13}{Department of Physics, Eidgenossiche Technische Hochschule (ETH), CH-8093 Zurich, Switzerland.}
\altaffiltext{14}{Institut de Recherche en Astrophysique et Plan\'etologie, CNRS, 14, avenue Edouard Belin, F-31400 Toulouse, France.}
\altaffiltext{15}{IRAP, Universit\'e de Toulouse, UPS-OMP, Toulouse, France.}
\altaffiltext{16}{INAF - IASF Milano, Milan, Italy.}
\altaffiltext{17}{Osservatorio Astrofisico di Arcetri, Largo Enrico Fermi 5, 50125 Firenze, Italy.}
\altaffiltext{18}{California Institute of Technology, MC 105-24, 1200 East California Boulevard, Pasadena, CA 91125.}
\altaffiltext{19}{National Optical Astronomy Observatory, 950 North Cherry Avenue, Tucson, AZ 85719, USA.}
\altaffiltext{20}{Space Telescope Science Institute, 3700 San Martin Drive, Baltimore, MD 21218, USA.}  
\altaffiltext{21}{Karl-Schwarzschild-Str. 1, Garching, D-85748, Germany.}
\altaffiltext{22}{Graduate School of Science and Engineering, Ehime University, Bunkyo-cho, Matsuyama 790-8577, Japan.}   
\altaffiltext{23}{Steward Observatory, University of Arizona, Tucson, AZ 85721, USA.}
\altaffiltext {24}{INAF - IASF Bologna, via Gobetti 101, 40129 Bologna Italy.}
\altaffiltext{25}{Institut d'Astrophysique de Paris, UMR 7095, CNRS, Universit\'e Pierre et Marie Curie, 98 bis Boulevard Arago, F-75014 Paris, France.}  
\altaffiltext{26}{Centro de Astronomia e Astrof'sica da Universidade de Lisboa, Observatorio Astronomico de Lisboa, Tapada da Ajuda, 1349-018 Lisboa, Portugal.}
\altaffiltext{27}{University of California, Department of Physiscs and Astronomy, Riverside, CA 92508, USA.}  
\altaffiltext{28}{The Hakubi Project, Kyoto University, Kitashirakawa-Oiwake-cho, Sakyo-ku, Kyoto 606-8502, Japan.}
\altaffiltext {29}{ASI  Science Data Center, via Galileo Galilei, 00044, Frascati,  Italy.}
\altaffiltext{30}{ de Astrof'sica de Andaluc'a, CSIC, Apartado de correos 3004, 18080 Granada, Spain.}
\altaffiltext{31}{Dipartimento di Astronomia, Universit\'a di Padova, Padova, Italy.}
\altaffiltext{32}{Institute for the Physics and Mathematics of the Universe (IPMU), University of Tokyo, Kashiwanoha 5-1-5, Kashiwa-shi, Chiba 277-8568, Japan.} 
\altaffiltext{33}{Research Center for Space and Cosmic Evolution, Ehime University, Bunkyo-cho, Matsuyama 790-8577, Japan.}
\altaffiltext{34}{University of California Observatories/Lick Observatory, University of California, Santa Cruz, CA 95064, USA}



\begin{abstract}
With this Êpaper, we release Êaccurate photometric redshifts Êfor 1692
counterparts to {\it ÊChandra} sources in the central Êsquare degree of
the ÊCOSMOS Êfield. ÊThe Êavailability Êof Êa Êlarge training Êset Êof
spectroscopic Êredshifts Êthat extends Êto Êfaint magnitudesÊenabled
photometric Êredshifts Êcomparable Êto Êthe Êhighest Êquality Êresults
presently Êavailable Ê for Ênormal Êgalaxies. Ê ÊWe Êdemonstrate Êthat
morphologically Ê extended, Êfaint Ê X-ray Êsources Ê without Êoptical
variability Êare more Êaccurately Êdescribed by Êa Êlibrary of Ênormal
galaxies Ê(corrected Êfor emission Êlines) Ê than byÊAGN-dominated
templates, Êeven if Êthese Êsources have AGN-like X-ray luminosities.
 Preselecting Êthe library on Êthe bases of the Êsource properties allowed us Êto reach an
accuracy $\sigma_{\Delta z/(1+z_{spec})}\sim 0.015$ with a fraction of
outliers Êof 5.8\% Êfor the Êentire {\it ÊChandra}-COSMOS Êsample. ÊIn
addition, Êwe Êrelease Êrevised Êphotometric redshifts Êfor Êthe Ê1683
optical Êcounterparts Êof the ÊXMM-detected Êsources Êover the Êentire
2\,deg$^2$ Êof ÊCOSMOS. Ê For Ê248 sources, Êour Êupdated Êphotometric
redshift differs Êfrom the previous Êrelease by $\Delta Êz>0.2$. These
changes are predominantly due to the inclusion of newly available deep
H-band Êphotometry (H$_{\rm AB}$=24\,mag). Ê We illustrate Êonce again
the Êimportance Êof Êa Ê spectroscopic Êtraining Êsample Êand Êhow Êan
assumption about the Ênature of a source together Êwith the number and
the Êdepth Êof the Êavailable Êbands Êinfluence Êthe accuracy Êof Êthe
photometric redshifts determined for AGN. ÊThese considerations should
be kept in mind when defining the observational strategies of upcoming
large Êsurveys Êtargeting ÊAGN, Êsuch Êas eROSITA ÊatÊX-ray  energies Êand
ASKAP/EMU in the radio band.
\end{abstract}






\section{Introduction}
The scientific yield of current and future systematic studies of large samples of extragalactic sources depends primarily on the observable redshift, which is one of very few observables that can be directly measured. A redshift then indicates the source distance via a cosmological model, and  can be used to estimate quantities such as age, black hole (BH) mass, and accretion rate. The constraint of source redshifts has been a primary goal of deep pencil-beam (e.g.,  HUDF: \cite{Williams:1996lr}), wide-area (e.g. AEGIS: \cite{Davis:2007uq}; COSMOS: \cite{Scoville:2007rw}, GOODS: \cite{Giavalisco:2004fj}; ECDFS: \cite{Lehmer:2005yq}; CFHTLS: \cite{Cuillandre:2006rt}), as well as future wide-field  synoptic sky surveys  across the  whole electromagnetic spectrum (e.g., {\it eROSITA}: \cite{Predehl:2007gf}; Pan-STARSS: \cite{Burgett:2009fr}; LSST: \cite{Ivezic:2006mz}; EMU: \cite{Norris:2010qf}; {\it WISE}: \cite{Duval:2004pd}).  Given the still limited number of multi-objects, near-infrared spectrographs  available on large telescopes,  comprehensive spectroscopic follow-up  studies are generally impractical for deep and large sky surveys and  the need for reliable photometric redshifts has arisen.

Thanks to the availability of extensive multi-waveband observations, the accuracy of the photometric redshifts of normal galaxies  has dramatically improved over the past decade. The main milestones  have been: the availability of deep near- and mid-infrared data for the surveys under study,  the use of intermediate-band filters  that help to increase the spectral resolution of the measured spectral energy distribution (SEDs) \citep[][]{Wolf:2001ul, Wolf:2003fk, Salvato:2009zw, Ilbert:2009hl, Cardamone:2010lr}), and the inclusion of emission  lines in the template SEDs of normal galaxies (Ilbert et al 2009; FORS Deep Field: Bender et al. 2001). As a result, we can now estimate the photometric redshifts of normal galaxies  with a 2\% accuracy \citep [see e.g.,][]{Ilbert:2009hl,Cardamone:2010lr}.

However, determining accurate and reliable photometric redshifts for sources dominated by an  active galactic nucleus (AGN) remain challenging for a  number of reasons. First of all, powerful  AGNs are dominated by a power-law SED, whose shape  produces a color-redshift degeneracy that only a complete and deep multi-wavelength coverage can break. Secondly, the galaxies that host an AGN contribute in most cases to the global SED of the source. The number of possible different types of galaxies and relative host/AGN contributions (as a function of wavelength) is so  large that degeneracies between templates and redshifts are unavoidable. Finally, flux variability is an intrinsic property of AGN  that many multi-wavelength surveys  do not take into account  when planning their observations, leading to problems in achieving a robust SED  fit.
Only when we correctly account for all these properties  will photometric redshifts (hereafter photo-z) for AGN become more reliable \citep[][]{Salvato:2009zw, Luo:2010qp, Cardamone:2010lr}.
 
\begin{figure}
\begin{center}
\includegraphics[scale=0.45]{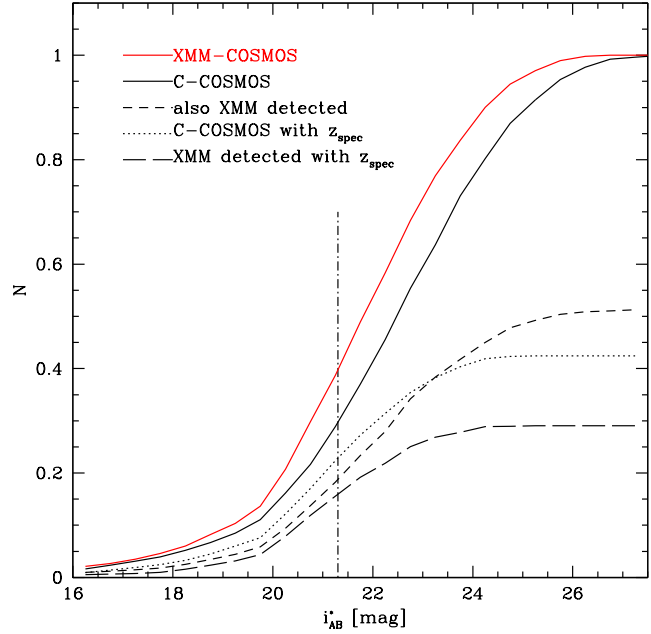}
\caption{ \small{Normalized cumulative {\it i}$^*_{AB}$ magnitude distribution for the optical counterparts of the {\it Chandra}-- (black solid line) and
 XMM-- (red solid line) COSMOS sources.  Distribution of sources common to both samples are also indicated (black dashed line). 
 The dotted curve indicates the C-COSMOS sub-sample  with  reliable spectroscopic redshifts, while the long-dashed curve  indicates the C-COSMOS sources with spectroscopic  redshifts and in common  with the XMM-COSMOS sample. The vertical line  represents the average magnitude of the spectroscopic sample available for C-COSMOS.}}
\label{fig:cumulate_mag}
\end{center}
\end{figure}

Expanding our previous studies \citep[][hereafter S09]{Salvato:2009zw} of the photo-z of the XMM observations of the entire 2 square degrees of the COSMOS field 
\citep{Hasinger:2007dn, Cappelluti:2009fj, Brusa:2010lr}, we provide photo-z for the counterparts  to $\approx 1700$ {\it Chandra} detected sources in the central 0.9 square  degrees \citep[][Civano et al., 2011, in preparation]{Elvis:2009kx, Puccetti:2009rt}.
 These {\it Chandra} data are significantly deeper (factor of $\sim$3...4)  than the XMM data, and their optical counterparts reach fainter magnitudes (Figure\,\ref{fig:cumulate_mag}). As a consequence, the  method developed to compute photo-z for the XMM-COSMOS sources  needs to be revised before its application to the {\it Chandra} data set.
 
 Our paper is structured as follows.  In \S\,\ref{sec:data}, we present the optical counterparts of our {\it Chandra} sources, in addition to our photometric and spectroscopic  analyses. In \S\,\ref{sec:preparation} we repeat the procedure introduced in S09 and split  the sample in two subsamples, on the base of the morphological and variability analysis. In \S\,\ref{sec:newPhotoZ}, we illustrate how we compute the photo-z, extending the technique to faint X-ray sources. We first compute the photo-z using exactly the same procedure used on S09, showing its limitations (\S\,\ref{subsec:asXMM}). We then discuss how the results can be improved in the following subsections.   In \S~\ref{sec:results} we presents our results, highlighting the properties of an individual source with z$_{phot} \sim$ 6.8 in  \S\,\ref{subsec:highz}.  General discussion and conclusions, using both {\it Chandra} and XMM sources are presented in  \S\,\ref{sec:discussion}, and \S\,\ref{sec:conclusions}, respectively. Throughout this  paper, we use AB magnitudes and assume  that H$_{0}$ =  70\,km s$^{-1}$ Mpc$^{-1}$, $\Omega_{\Lambda}=0.7$, and $\Omega_{M}=0.3$. 
  
  
\section{The {\it Chandra} COSMOS sample}
 \label{sec:data}

\subsection{Optical and near-IR counterparts}

The Chandra COSMOS  survey (hereafter C-COSMOS; Elvis et al. 2009) is a large (1.8 Ms) {\it Chandra} program covering the central 0.5\,deg$^2$ of the COSMOS field (centered at 10\,h  , +02\degs )  with an effective exposure of $\sim 160$ ks, and an outer 0.4\,deg$^2$ area with an effective exposure of $\sim$80 ks. The limiting depths  of the point-source detections are $1.9\times10^{-16}$ erg  cm$^{-2}$ s$^{-1}$ in the soft ($0.5-2$\,keV) band, $7.3 \times10^{-16}$\,erg cm$^{-2}$ s$^{-2}$ in the hard ($2-10$ keV) band, and $5.7\times10^{-16}$\,erg cm$^{-2}$ s$^{-1}$ in the full ($0.5-10$ keV) band.

A total of 1761 X-ray point sources  were detected  in our {\it Chandra}  data \citep[for details on the source detection procedure see][]{Puccetti:2009rt}.
The X-ray catalog  was presented in \cite{Elvis:2009kx}. The  optical/NIR counterparts were identified on the basis of a likelihood ratio technique (ML) applied to our optical \citep{Capak:2007db}, near-infrared \citep{McCracken:2010pj}, and Spitzer/IRAC \citep{Sanders:2007qd, Ilbert:2010qy} catalogs, and are presented in Civano et al. (2011, in preparation), together with an overall analysis of the sample properties. In summary, thanks to this multi-wavelength identification approach, 1753  counterparts to our X-ray sources (i.e. 99.6\%) have been successfully identified in optical/IR bands. Of these 1753, 42 are nearby stars or sources  that are too close to a star to be detected separately; these stellar sources are not considered in this paper. 
 
For completeness, we provide in Figure\,\ref{fig:cumulate_mag}, the normalized cumulative {\it i}$^*_{AB}$ magnitude distribution for the optical counterparts  to C-COSMOS (black solid line) compared to the distribution for XMM-COSMOS (red solid line). The distribution of sources common to both samples (black short dashed line) and the distribution of sources with available spectroscopic  redshifts (see more in \S\,\ref{sec:newPhotoZ}) are also indicated (dotted and long dashed lines). 
 
Here we present the photo-z  of the 1692 sources for which a large number (15$\ge \rm N_{filters} \le 31$) of reliable photometric data are available. 1677  of these sources have an optical counterpart in the updated, publicly available photometric catalog down to {\it i}$^*_{AB}$=26.5 mag\footnote{\url{http://irsa.ipac.caltech.edu/data/COSMOS/tables/photometry/}. This catalog includes the photometry in all the 25 optical/NIR broad-, intermediate- and narrow-bands filters,  from ``u'' to ``Ks'' . the photometry is  computed at the position of the  {\it i}$^*$-band image, using Sextractor \citep{Bertin:1996qy} in dual mode. The catalog supersedes Capak et al. (2007), with improved source detection and photometry extracted in 3" apertures}.
 
An additional 15 objects were found  in the K-band catalog (McCracken et al. 2010) and aperture photometry were extracted in all broad-band optical/near-IR and COSMOS bands using the K-band images as reference.  We note that 11 of these 15 sources are also clearly visible in the optical images but are not present in the updated optical catalog  because they are either close to a saturated source or  below the detection limit.  \\

For these 1692 sources, a coordinate cross-match (up to 0.5'') was performed between the optical catalog and  the {\it Spitzer}/IRAC \citep{Sanders:2007qd} and {\it GALEX} \citep{Zamojski:2007lq} catalogs.
To create the {\it GALEX} catalog, the $U$-band image was used as a prior, via PSF fitting. 
Thus, the risk of wrong optical/UV identification is much lower. 
The  IRAC images are deep ([3.6 $\nu\,m$]$_{\rm AB} \sim $24 mag) but they have  large  PSF. 
We performed simulations which have shown
that not  more that 10\% of  the photometry of the  $\sim$ 400,000 sources of
the  IRAC catalog  may be  effected  by blending.  
This is not effecting the  associations optical/IRAC
as we visually inspected the associations \citep[][Civano et al. 2011, in preparation]{Brusa:2010lr}. However, the blending can 
effect the photometry of few sources, explaining the origin of a small number of outliers (see also discussion in S09).

An additional 19 sources are neither detected in our optical images, nor listed in the K-band catalog, but  have a clear counterpart  in the 3.6$\mu$m images. Although these sources are potentially at high redshift, we do not  attempt to estimate their photo-z as they have the same properties as the sources presented in \S\, 5.3 of S09 (in  nine cases they are actually the same sources).
There, the formal best-fit redshift was shown to be higher than 4, but the redshift probability distribution function (PDFz) indicated that there were  insufficient constraints to reject a solution at lower redshift. For these sources, only deeper photometry could provide reliable constraints and photo-z.

\section{Morphological and variability analysis} 
\label{sec:preparation}
 
In S09, the optical counterparts of the {\it XMM} sources \citep[presented in][]{Brusa:2010lr} were
divided into two sub-samples depending on their morphological and temporal properties. Objects that appeared as point sources \citep[as defined in][]{Leauthaud:2007fj} in deep COSMOS HST/ACS images \citep{Koekemoer:2007kx} and/or showed  brightness fluctuations were grouped as {\it QSOV} (short for point-like or varying) and their photometry was corrected for variability, if necessary. For this purpose, we introduced a parameter, VAR (Eq. 1 in S09), that describes the deviation of the optical photometry  from a reference epoch (2006, time of quasi-simultaneous  optical and {\it Spitzer}/IRAC  observations).
On the basis of the distribution of this parameter for the entire  XMM-COSMOS sample,
the photometry for the sources with $VAR>$0.25 mag were corrected. The threshold was chosen as the  value of  $VAR$ at which the sample of extended sources in XMM displayed a  sharp decline distribution  (Figure 1 in S09). Sources that were not grouped in the {\it QSOV} sample were then classified as extended and non-varying ($VAR<0.25$) and were assigned to the {\it EXTNV} group. The identification of these  two subgroups permitted us to use a luminosity prior  that is typical  of AGN for the {\it QSOV} sample (see \S \ref{sec:qsov}), which in turn reduced the  parameter space of the possible photo-z solutions and thus the degeneracies.

\begin{figure}
\begin{center}
\includegraphics[scale=0.45]{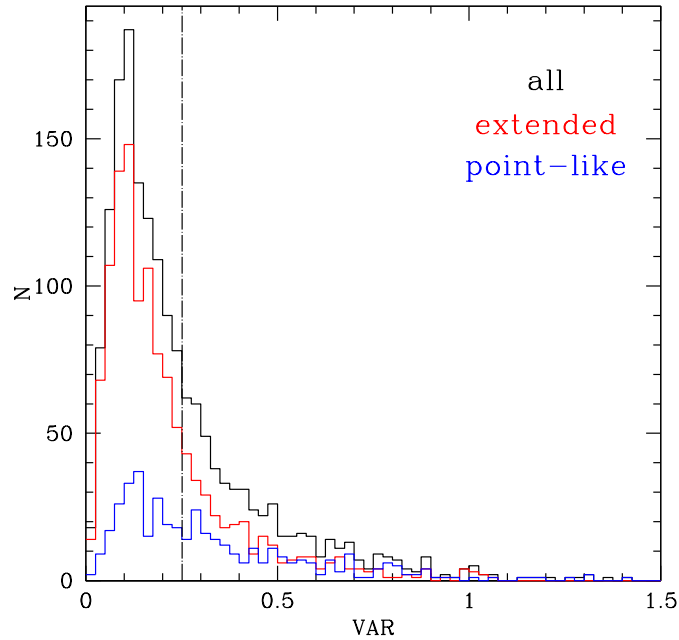}
\caption{ \small{VAR histogram distribution for extended and point-like sources. As for XMM-COSMOS, we adopted the value VAR=0.25 as  a threshold beyond which we correct the photometry for variability.}}
\label{fig:histomag}
\end{center}
\end{figure}

The C-COSMOS sample discussed in the following  was treated in an identical manner, and its $VAR$ distribution is shown in Figure\,\ref{fig:histomag}. Compared to the XMM-COSMOS sample, the VAR distribution does not show a drop at VAR=0.25 mag,  This is expected, due to the deeper observations of the  smaller area of C-COSMOS. However, we decided to adopt the same value in order to limit as much as possible the number of caveats and  allow a more general procedure to be adopted.
 
We find that 766 sources satisfy the criteria for the {\it QSOV} sample, while 926 are classified as {\it EXTNV} sources. Among the {\it QSOV} sources, 442 (58\,\%) were already included in the  XMM sample, while the {\it EXTNV} sample contains 421 sources (46\,\%), which were also
detected with XMM.
 

\section{Photometric redshift }
\label{sec:newPhotoZ}

In the  following subsections we describe the  photo-z technique used
for the C-COSMOS  sources.  As in S09, we  used the publicly available
{\it              Le             Phare}             code\footnote{\url
{http://www.oamp.fr/people/arnouts/LE$\_$PHARE.html}}  \citep{Arnouts:1999lr,  Ilbert:2006vl},  which is based on a  $\chi^2$ template-fitting procedure.
The templates that we used  were either used for computing the photo-z
for  normal galaxies in  I09, or  used for  computing the  photo-z for
XMM-COSMOS in S09. The I09 templates include elliptical and spiral galaxy templates from  \cite{Polletta:2007cs} They also include blue star-forming galaxies generated with  \cite{Bruzual:2003uq}.
The S09 templates include some of  the AGN library from \cite{Polletta:2007cs}, and hybrid templates combining AGN and normal galaxies. 
 How we  created the  templates,  how we  settled on  the
libraries,  and  how they  compare  with  other  libraries are  widely
described in I09 and S09, respectively. Extinction is added to the templates as a free parameter in the fit. We used the  \cite{Calzetti:2000rt}  and the \cite{Prevot:1984bf} attenuation laws. We also calibrated the zero-points of the photometric catalogue using the spectroscopic redshift sample of normal galaxies, as described in \cite[][I09]{Ilbert:2006vl}.  We did not  allow any galaxy to be brighter than M$_{\rm B}$=-24. For AGN, the luminosity prior is more complex and depends on the classification {\it EXTNV}/{\it QSOV} (see \S \,\ref{subsec:asXMM}). Finally, the full redshift probability  distribution function is also derived.

After estimating the photo-z we assessed the accuracy by comparing our results with  712 (21) reliable spectroscopic redshifts of galaxies (stars). \\
The spectroscopic redshifts were either publicly available via SDSS (DR8) or  obtained within the COSMOS collaboration. In fact, the counterpart of X-ray targets were the primary targets of Magellan/IMACS \citep{Trump:2007fv} and MMT \citep{Prescott:2006ek} campaigns, or secondary targets in the zCOSMOS and zCOSMOS-deep surveys at VLT/VIMOS \citep[][Lilly et al. 2011, in preparation]{Lilly:2007qr,Lilly:2009qy}, or again obtained at  Keck/DEIMOS (PIs: Scoville, Capak, Salvato, Sanders, Kartaltepe) and FLWO/FAST \citep[][]{Wright:2010qy}, respectively. 
While the spectroscopic sample used for the training for XMM-COSMOS reached a luminosity of $<{\it i}^*_{AB}>$=22.5 mag, the new sample  reaches magnitudes of {\it i}$^*_{AB}$=25.4 mag ( $<{\it i}^*_{AB}>$=21.3 mag; vertical dotted-dashed line in Figure ~\ref{fig:cumulate_mag}), thus  provides some insight into the faint source population. It is important to stress that {\it all} the spectroscopic redshifts have an probability higher than 75\% to be secure, as  at least  two emission/absorption features were used for the redshift determination. 

 Throughout the paper, we measure the accuracy of the photo-z using the normalized median absolute deviation \citep[NMAD,][]{Hoaglin:1983bh} defined as 
$\sigma_{NMAD}$=1.48 $\times$ median($|$z$_{\rm phot}$- z$_{\rm spec}|$/(1 + z$_{\rm spec}$)) 
For a gaussian distribution, $\sigma_{NMAD}$ is directly comparable to  the definition adopted in other papers that directly quote the $\sigma_{\rm \Delta z/(1 + \rm z_{\rm spec})}$.
This dispersion estimate is relatively insensitive to catastrophic outliers (i.e., objects with $|$z$_{\rm phot}$ Ð z$_{\rm spec}|$/(1 + z$_{\rm spec})>$0.15). 
The fraction of outliers is denoted by $\eta$.
After applying a method identical to that used for the XMM sample, we discuss how to improve the reliability of photo-z for the {\it EXTNV} and {\it QSOV} sub-samples, respectively.


\subsection{Estimating C-COSMOS photo-z as for XMM-COSMOS}
\label{subsec:asXMM}

To understand whether or not C-COSMOS  is sampling the same population  as XMM-COSMOS sources, we first computed photo-z following the same procedure as described in detail in S09.
In particular, after dividing the  C-COSMOS sources into {\it EXTNV} and {\it QSOV}, we used the same  template library, consisting mostly of AGN and  hybrid templates. The  hybrid templates are constructed by combining galaxy and AGN empirical SEDs (details   of the templates and  their construction is fully described in S09).
Furthermore, the same luminosity prior on the absolute magnitude in $B$-band ($-20 >M_B> -30$)  was applied to the {\it QSOV} sample.
 
We compare the resulting photo-z with the spectroscopic sample of 712 sources. It is important to stress that the spectroscopic sample is a close approximation to a {\it blind} sample, different in his properties from the sample used as  {\it training} of the photo-z for XMM-COSMOS.  Indeed, only 273 of these sources were included in the original training sample used in S09. The new 439 sources had either  no spectroscopy at that time, or lie below the XMM flux limit.

\begin{figure}[ht]
\begin{center}
\includegraphics[scale=0.40]{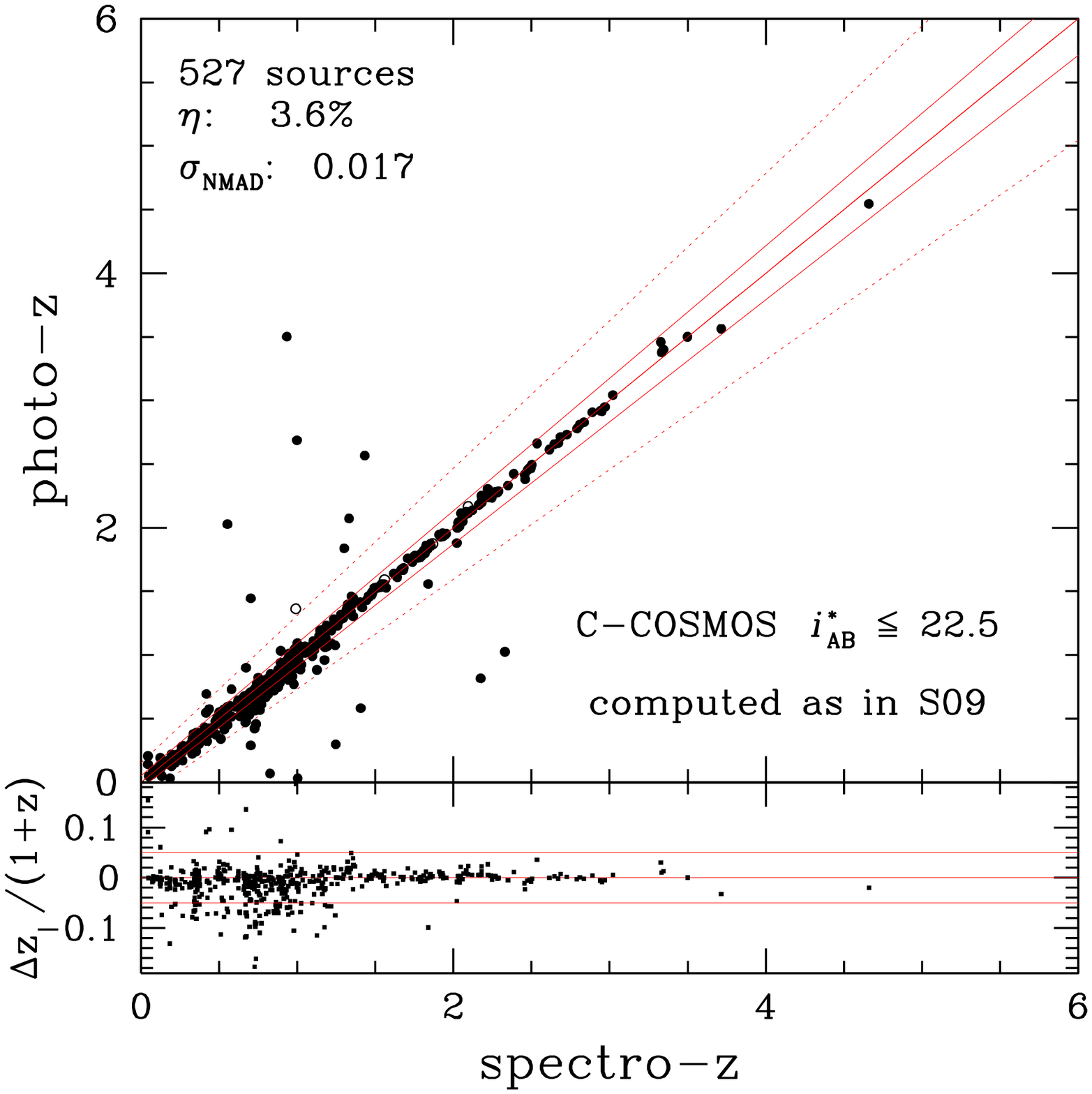}  
\includegraphics[scale=0.40]{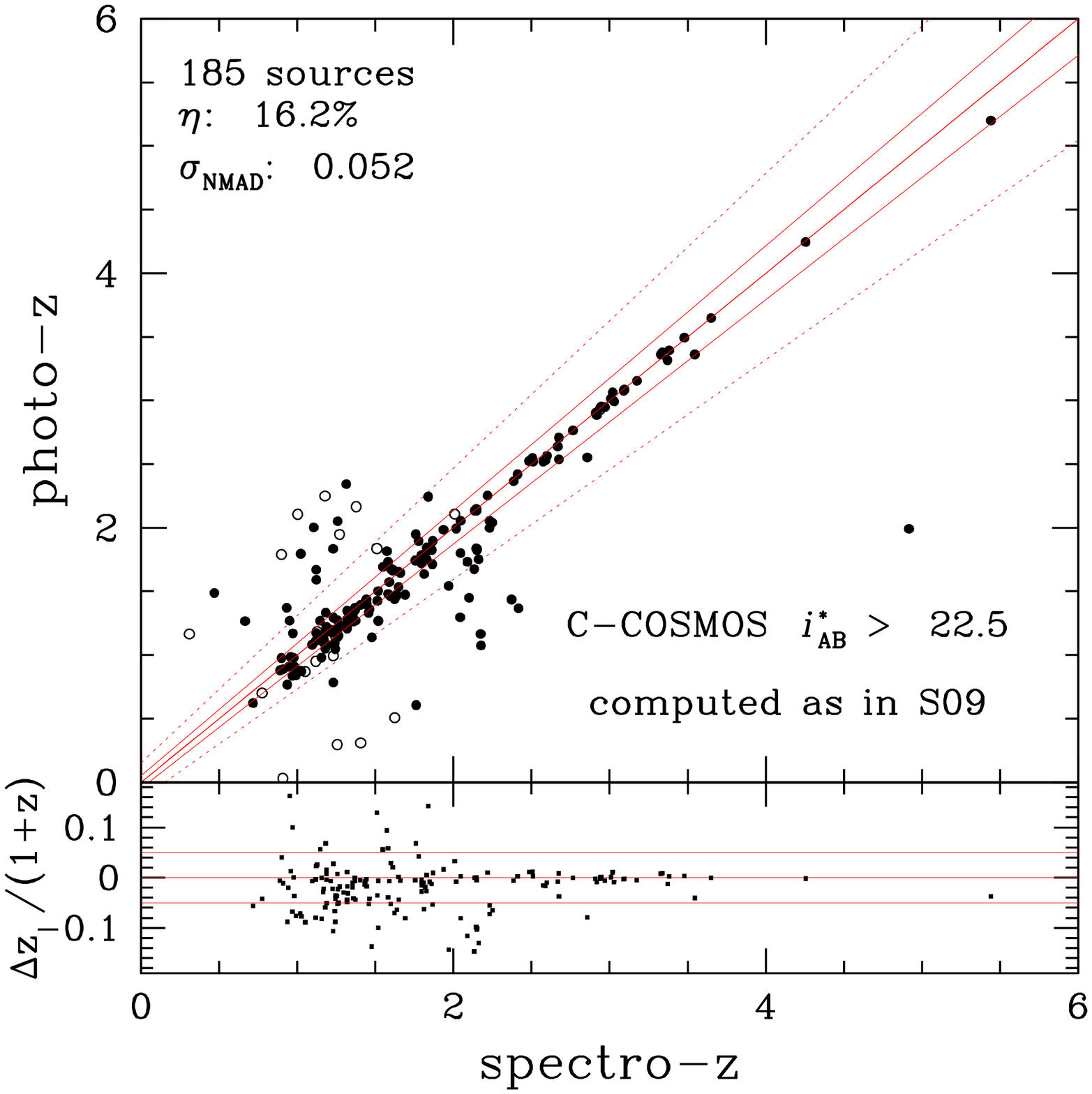}
\caption{ \small C-COSMOS photometric redshifts computed following the recipe defined in S09, compared to the spectroscopic redshifts. The comparison is shown  for sources brighter (left panel) and fainter (right panel) than {\it i}$^*_{AB}$=22.5 mag. Open circles represent sources for which there is at least a second significant peak in the redshift probability distribution. The solid lines correspond to z$_{\rm phot}$=z$_{\rm spec}$ and  z$_{\rm phot}=\pm$0.05(1+\rm z$_{\rm spec}$), respectively. The dotted lines limit the locus where z$_{\rm phot}$=$\pm$0.15(1+z$_{\rm spec}$). While the quality of the photo-z for the bright sample  is comparable to the one obtained for the XMM-COSMOS sources without any new tuning or training, the photo-z computed for the fainter sources are significantly worse in terms of both dispersion and fraction of outliers.}
\label{fig:asXMM}
\end{center}
\end{figure}

While most of the photo-z are still excellent, the resulting fraction of outliers ($\eta$=9.0\%)  and accuracy ($\sigma_{NMAD} \sim$0.031) do not reach the quality obtained for the XMM-COSMOS sample ($\eta$=5\%, $\sigma_{NMAD}$=0.015 for sources {\it i}$^*_{AB}<$24.5\,mag). In particular, if we consider only the C-COSMOS sources brighter than {\it i}$^*_{AB}<$22.5\,mag (limit of the spectroscopic training sample used in XMM-COSMOS) the accuracy for the {\it EXNV} and {\it QSOV} sub-samples is the same as for XMM-COSMOS, even if the spectroscopic sample used for the comparison is not the same.
In contrast, for  sources fainter than {\it i}$^*_{AB}=$22.5\,mag we found a significant increase  in the fraction of outliers and  lower accuracy (compare Figure\,\ref{fig:asXMM} of this paper with Figures~4+12 in S09) are obtained. 

The comparable quality of the photo-z between C-COSMOS and XMM-COMOS at {\it i}$^*_{AB}<$22.5\,mag suggests that the optically bright populations probed by XMM and {\it Chandra} are similar and that the template library used in S09 is largely representative of their properties. In S09, there was no spectroscopic training sample for {\it i}$^*_{AB}>$22.5\,mag and the quality assessment for faint sources ({\it i}$^*_{AB}>$22.5\,mag) was based on the comparison with only 46 spectroscopic redshifts. The faint C-COSMOS spectroscopic sample now includes a total of 185 sources with  i$^*_{AB}>$22.5\,mag and the lower photo-z quality may  indicate that a new treatment, different from that used for the bright sample, is required.

 
\subsection{Revised treatment for the C-COSMOS {\it EXTNV} sample}
 \label{sec:extnv}

Twenty-four out of the 30 templates used to compute the photo-z for the XMM-COSMOS sources are dominated (from the 10\% to 100\% level) by an AGN component. On the other hand, as C-COSMOS extends to faint X-ray sources and thus also to faint and potentially optically obscured sources, one could argue that the library used  to analyze the XMM data is not fully representative.   It might be beneficial to consider a library including a set of ``pure galaxy'' templates.
This is particularly true for the {\it EXTNV} sub-sample, which contains predominantly nearby sources where the optical/near-IR emission is expected to be dominated by the host galaxy light. 

To assess the impact of a different library of templates, we computed the photo-z using the library and settings defined in I09. These authors used a library of 31 templates of normal galaxies to compute the photo-z  of two million normal galaxies ({\it i}$^*_{AB}<$26.5) in the entire COSMOS field, reaching an accuracy of $\sigma_{NMAD} \sim$ 0.015 with a fraction of outliers $\eta <5\%$. In particular, the authors included emission lines in the templates, as they were shown to contribute to various colors by up to 0.4 mag.

In Table\, \ref{tab:extnv}, we compare the resulting  quality of the photo-z for bright ({\it i}$^*_{AB}<$22.5) and faint ({\it i}$^*_{AB}>$22.5) {\it EXTNV} subsamples with the results obtained using the S09 library. 

From the comparison of the dispersions obtained in the bright, faint and full optical ranges it seems that a library of normal galaxies works generally better for the {\it EXTNV} sample than a library which includes AGN templates.  
However, for the bright sample, the fraction of outliers obtained  when using the library of normal galaxies is almost twice  that obtained using the library of AGN-dominated templates, indicating the need  for the former library for at least some sources. In addition, the result is consistent with what is found for the  sources in XMM-COSMOS, after recomputing the photo-z using now also the H- band photometry (last columns of \ref{tab:extnv}).

\begin{figure*}
\begin{center}
\includegraphics[scale=0.4]{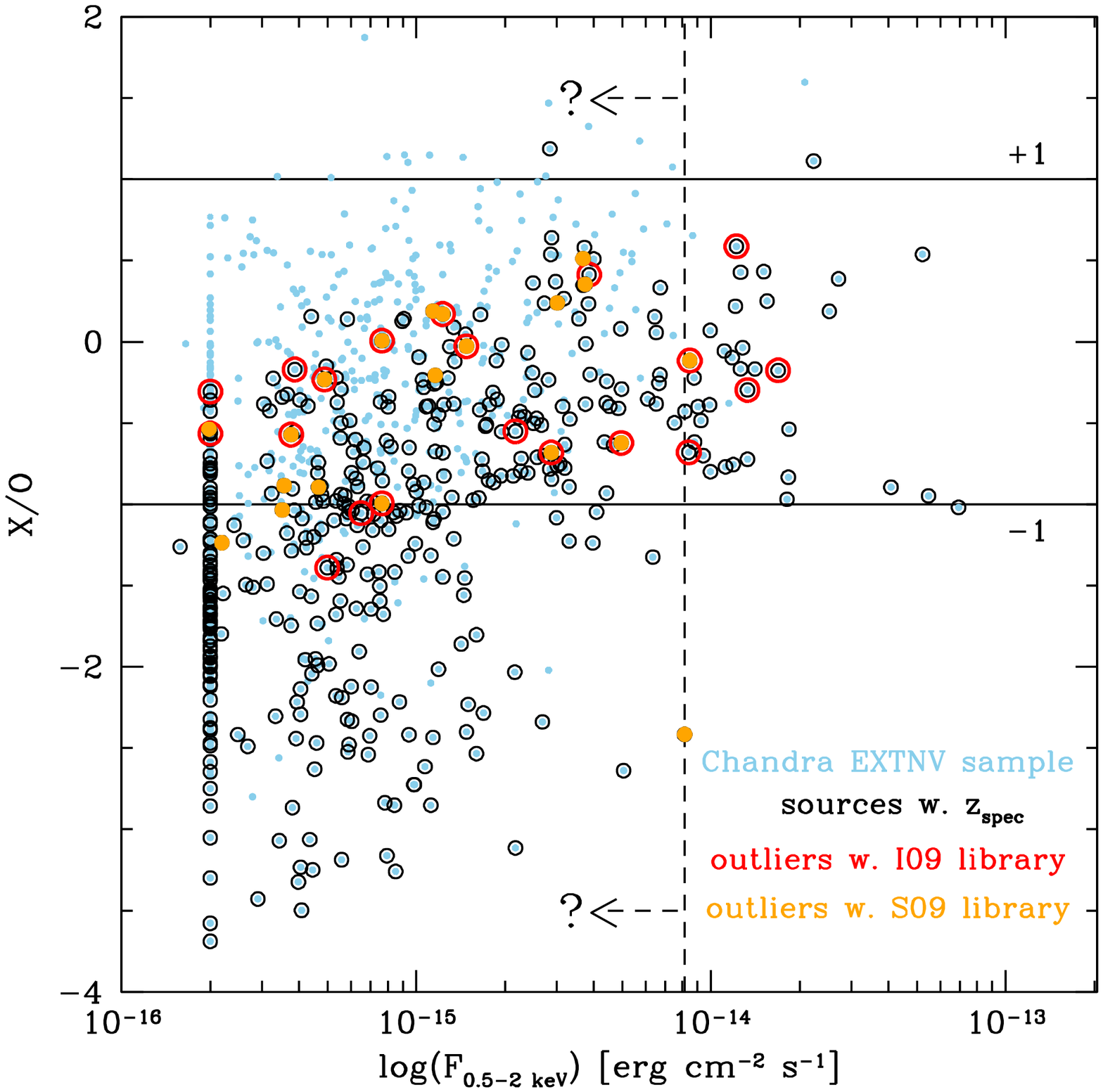}
\includegraphics[scale=0.4]{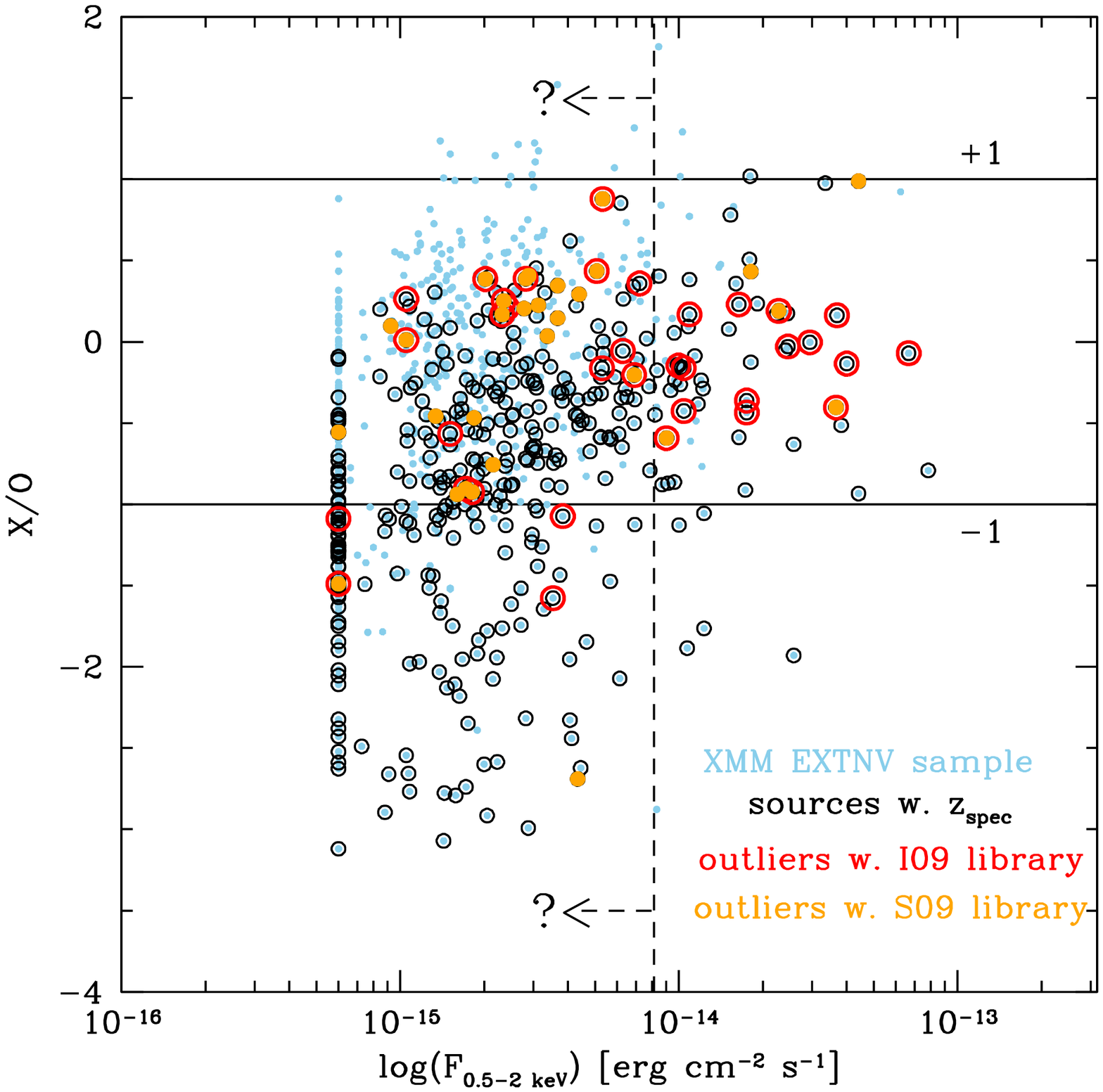}
\caption{Distribution of outliers for the {\it EXTNV} samples  (Left: C-COSMOS; Right: XMM-COSMOS) as a function of X/O and soft X-ray flux and compared to the rest of the sources distribution.  Light blue dots represent all the sources, while black circles represent sources with spectroscopic  redshift. Red circles indicate outliers for the library of normal galaxies of I09, while yellow filled circles indicate the outliers for the AGN-dominated S09 library. The distribution of  outliers is the same along the X/O axis. However, for each library the   outlier fraction depends on the X-ray flux of the source. While there is an  excess of outliers at bright X-ray fluxes for normal galaxy templates, the inverse occurs for the library of AGN-dominated templates at the faint end of  X-ray fluxes.}  
\label{fig:extnv}
\end{center}
\end{figure*}

To characterize the outliers and see whether they depend on the properties of the sources, we plot in the left panel of Figure\,\ref{fig:extnv} all the {\it EXTNV} sources as a function of their soft X-ray flux and their X/O ratio  \citep[][]{Maccacaro:1988lr}. In this specific case, using the soft X-ray flux and the optical ${\it i}^*$ AB magnitude, the X/O ratio is defined as log(F$_X$/F$_{opt}$)=log(F$_{(0.5-2 keV)}$+5.57+${\it i}^*$ [AB]/2.5. The ratio can be used as  a first order assessment of the nature of a source, with a galaxy  been characterized  by X/O$<$-1.5 and  an AGN-dominated source as -1$<$X/O$<$ 1.

 Both libraries are clearly able to reproduce the spectroscopic sample of galaxy-dominated sources  because in the range X/O$<$-1.5 there are virtually no outliers.
In addition within the locus of  AGN-dominated sources, the distribution of outliers  when using either library (red open circles and yellow filled circles for I09 and S09, respectively) is independent  of the X/O ratio.
The only real difference is visible in the distribution of outliers as a function of X-ray flux, where the library of AGN-dominated templates  provides more reliable photo-z at high X-ray fluxes, with only 2 outliers above F$_{(\rm 0.5-2 keV)}>8 \times 10^{-15} {\rm erg}/{\rm cm^2}/{\rm s}$ in contrast to the 5 for the library of normal galaxies. This is consistent with the fact that the extended, optically bright and X-ray bright sources in our sample are nearby (z$<$1) Seyfert or QSO. Indeed, all the sources with spectroscopic redshift and with   F$_{(\rm 0.5-2 keV)}>8 \times 10^{-15} {\rm erg}/{\rm cm^2}/{\rm s}$ have an absolute $B$ magnitude M$_{\rm B}<-20$ which is typical for AGN \citep[e.g.][]{Veron-Cetty:1998lr}. 
In contrast, fainter X-ray sources are either host dominated or low luminosity or obscured AGN for which the templates of normal galaxies are able to mimic the SED, thus correctly reproduce the redshift.

 On the basis of the available spectroscopic sample (open black circles), we argue that adopting a threshold at F$_{(\rm 0.5-2 keV)}>8 \times 10^{-15} {\rm erg}/{\rm cm^2}/{\rm s}$ and using the library of either normal galaxies or AGN-dominated templates for sources, respectively, below or above this value,  improves the accuracy of the photo-z, as demonstrated in the last row of Table\,\ref{tab:extnv} (indicated as ``combined''). 
 
\begin{figure*}
\begin{center}
\includegraphics[scale=0.40]{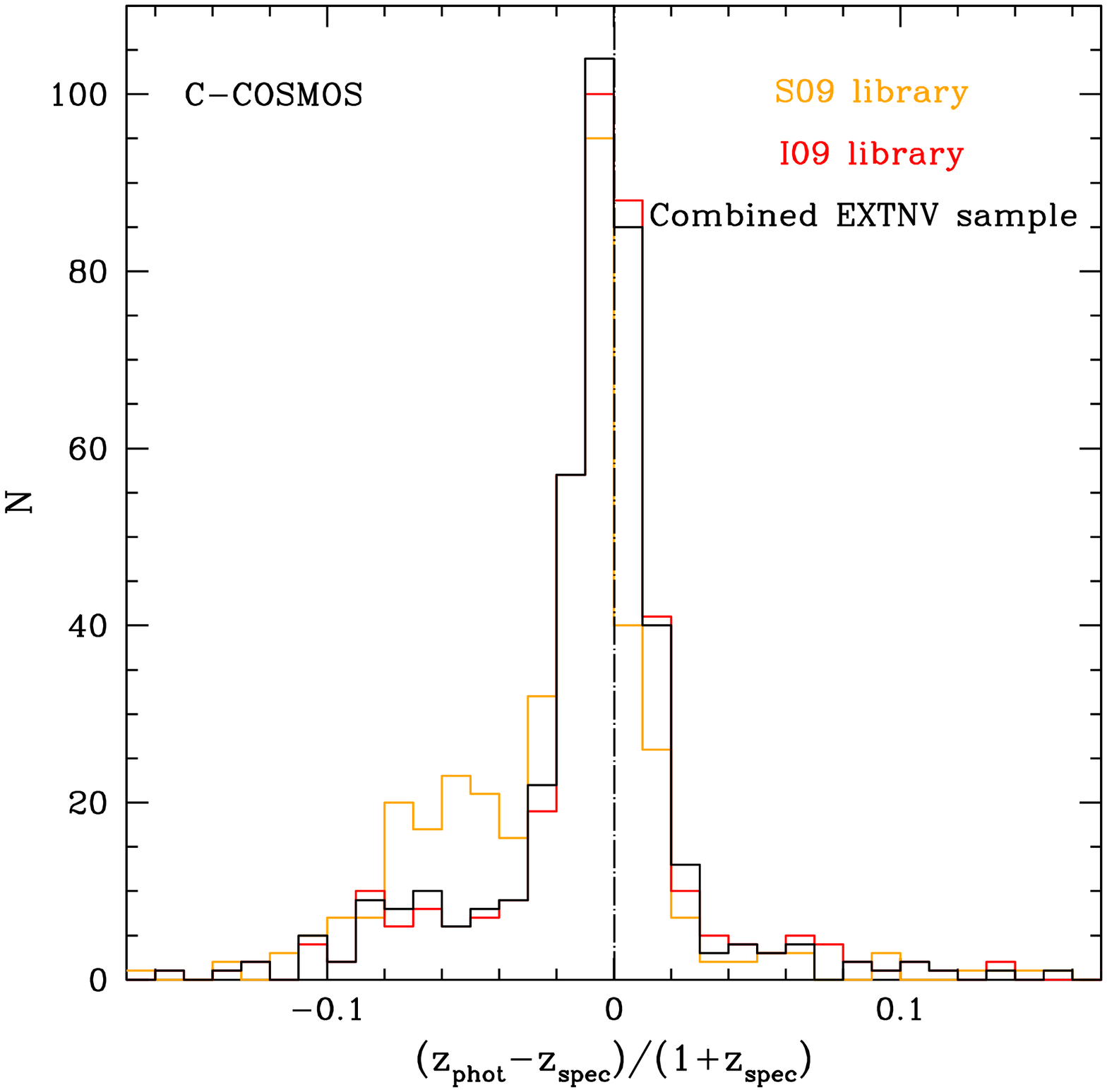}
\includegraphics[scale=0.40]{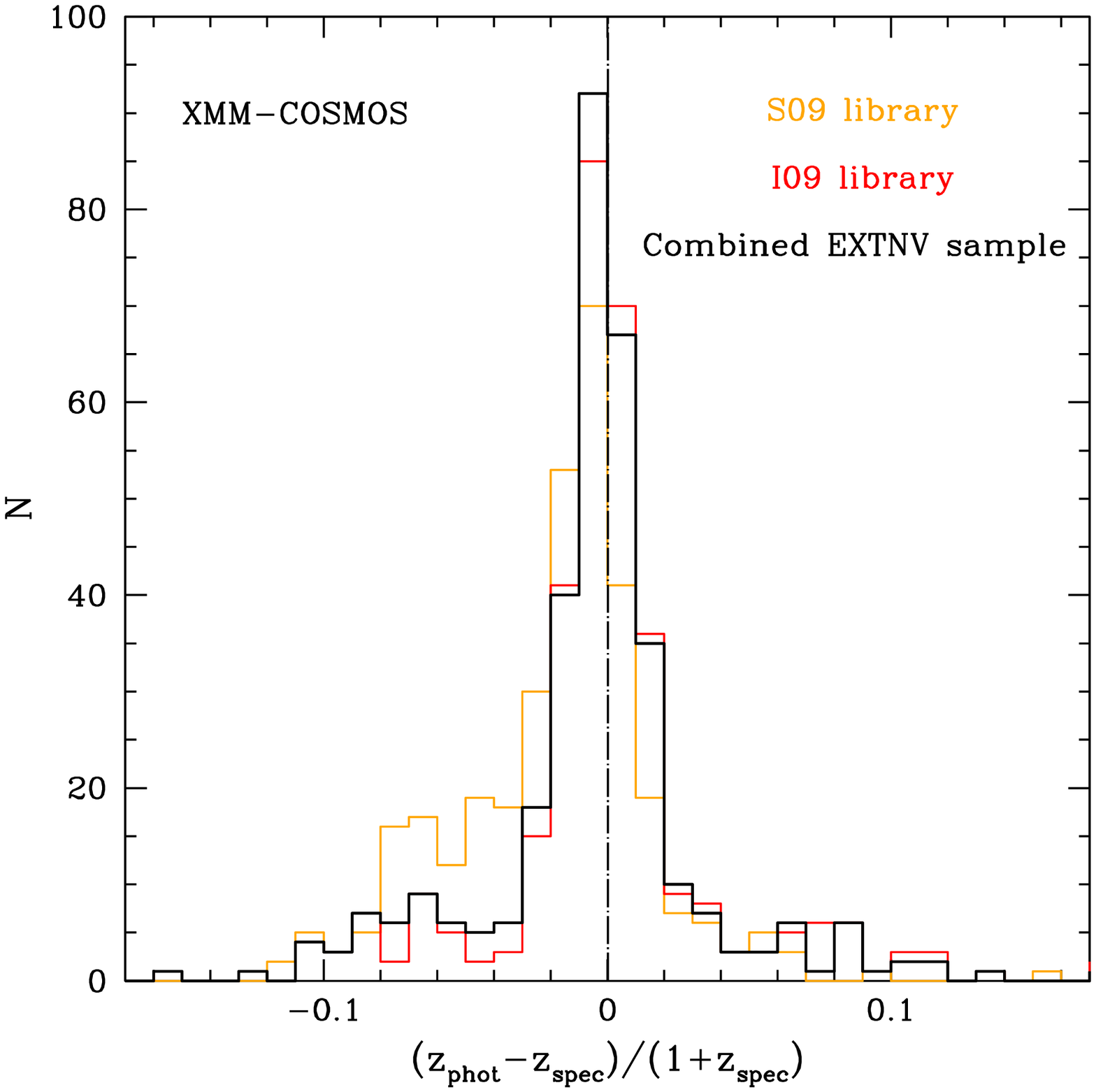}  
\caption{Left:$\Delta z/(1+z_{\rm spec}$) distribution for C-COSMOS {\it EXTNV} sub-sample using the I09 library of normal galaxies (red solid line), and the S09 library of AGN (yellow solid line). 
The black solid line  indicates the final result using S09 for F$_{0.5-2 \rm keV} < 8 \times 10^{-15} {\rm erg\,cm^{-2} s^{-1}}$ and I09 for fainter sources. 
Right: the same but for XMM-COSMOS sample.}
 \label{fig:histo_dz}
\end{center}
\end{figure*}

Given the  small number of outliers in the bright end of the X-ray flux, one could argue that the introduction of different templates depending on the X-ray flux is  unnecessary and that for the {\it EXTNV} the library of normal galaxies could be used by default. However,  for wide-field shallower X-ray surveys such as XMM-COSMOS, where a large number of bright X-ray sources are detected,  the use of AGN-dominated templates in the library  is more important (see right panel of Figure\,\ref{fig:extnv}). This new approach allows  us to reduce the fraction of outliers at any X-ray flux and, at the same time,  reduces the dispersion, which  is now symmetric and peaks at $\Delta {\rm z/}(1+\rm z_{spec})$=0 (compare yellow and black histograms in Figure\, \ref{fig:histo_dz} for C-COSMOS and XMM-COSMOS, respectively).

We note that the adopted X-ray threshold is chosen to minimize the number of outliers and thus it strongly depends on the spectroscopic sample available for the comparison. The value  is, at the moment, fixed where the first outlier for the AGN-dominated library  appears in the C-COSMOS {\it EXTNV} sample, but  could possibly be moved  to fainter X-ray fluxes, depending on the availability of future spectroscopy in the range F$_{\rm 0.5-2 keV}= 4...8\times 10^{-15} {\rm erg}/{\rm cm^2}/{\rm s}$. 
 

\subsection{Analysis of the {\it QSOV} sample}
\label{sec:qsov}

\begin{figure}
\begin{center}
\includegraphics[scale=0.40]{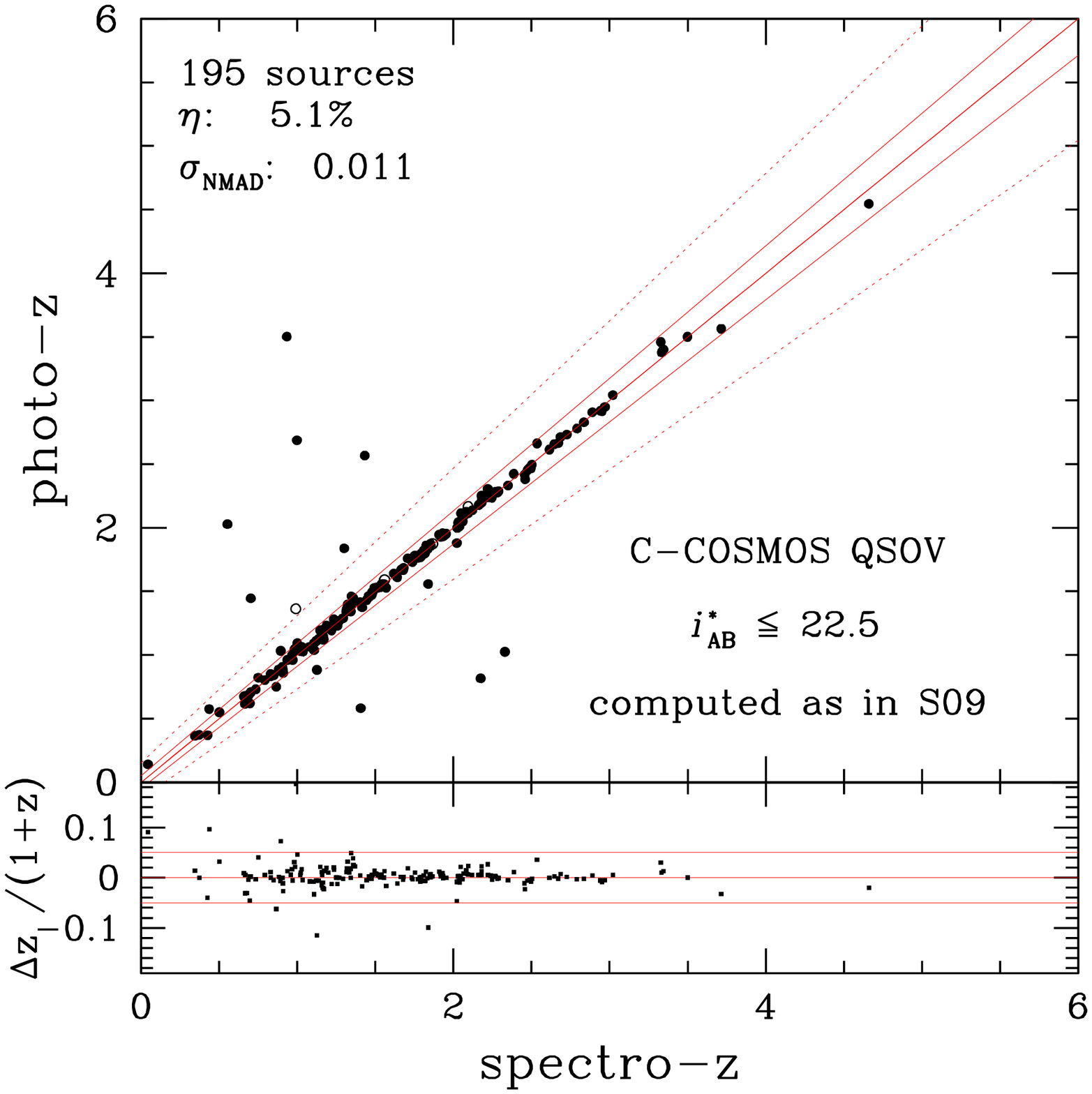} 
\includegraphics[scale=0.40]{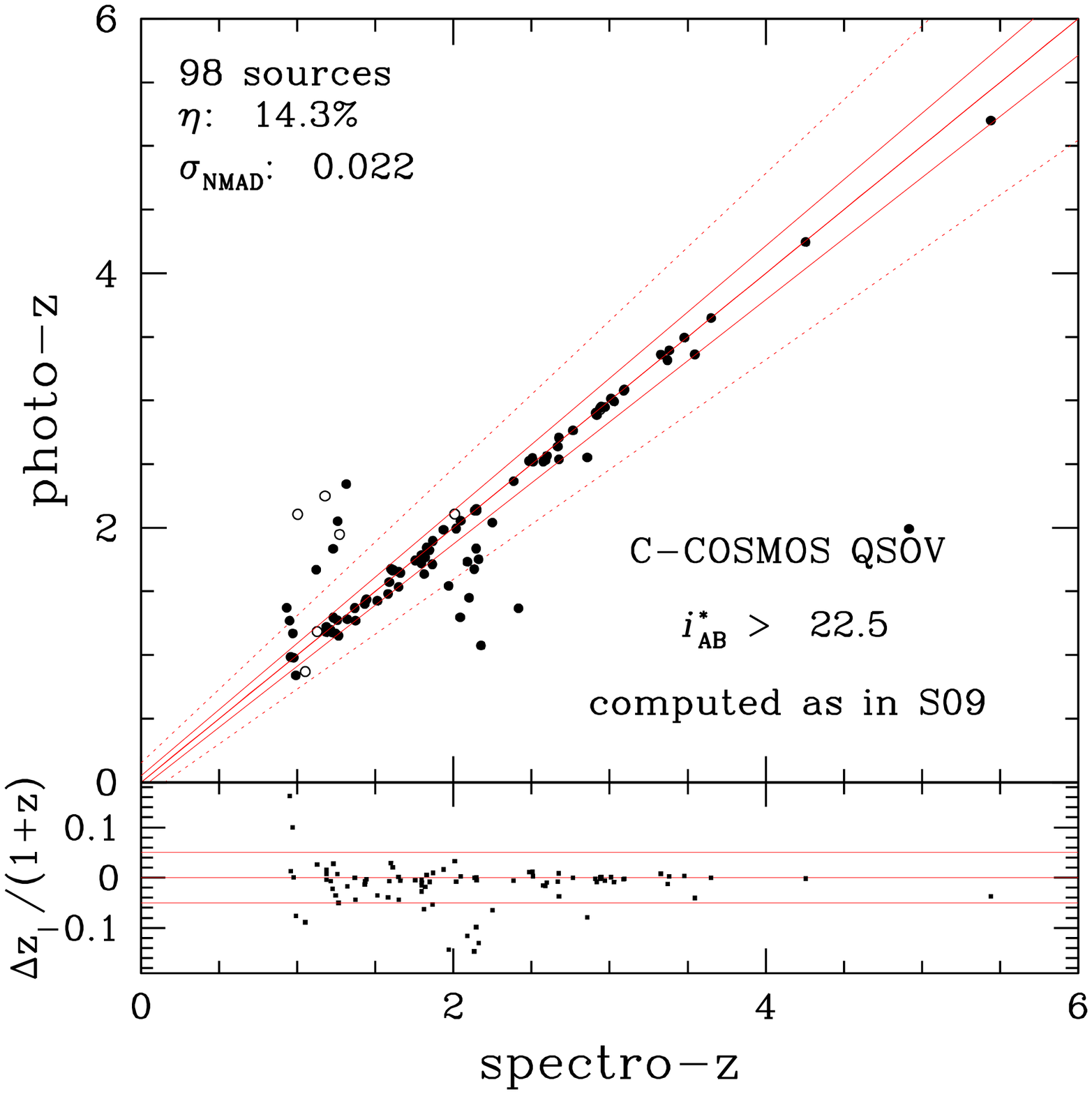} 
\caption{\small Comparison between spectroscopic  and  photo-z computed as in S09 for the C-COSMOS {\it QSOV} sources, brighter (left panel) and fainter (right panel) than {\it i}$^*_{AB}$=22.5. Black open circles indicate sources with a second possible solution in the redshift probability distribution.
Again, the quality of the photo-z for the bright sample is comparable to  that obtained for the XMM-COSMOS {\it QSOV} sources without any  additional tuning, even if the spectroscopic training sample is different.}
\label{fig:zpzs_CC}
\end{center}
\end{figure}

 As for the C-COSMOS {\it EXTNV} sample, the photo-z accuracy  of C-COSMOS {\it QSOV} sources is identical to that achieved for XMM {\it QSOV} sources when the analysis is limited to sources brighter than {\it  i}$^*_{AB}=22.5$\,mag ($\sigma_{NMAD}=0.011$ and $\eta=5.1$\,\%).
However, the fraction of outliers  increases to $\eta=14.3$\,\% (consequently, $\sigma_{NMAD}$=0.22) if we limit the analysis only to the 98 sources fainter than {\it  i}$^*_{AB}=22.5$\,mag.
We find indeed that $\sim 60\%$ of the outliers in the {\it QSOV} sample is represented by faint sources (Figure\,\ref{fig:zpzs_CC}) and that  they are concentrated in two redshift ranges, where the photo-z are systematically overestimated (1$<z_{\rm spec}<$1.5) or underestimated ( (2$<z_{\rm spec}<$2.5)  relative to the spectroscopic redshifts.

\begin{figure}
\begin{center}
\includegraphics[scale=0.40]{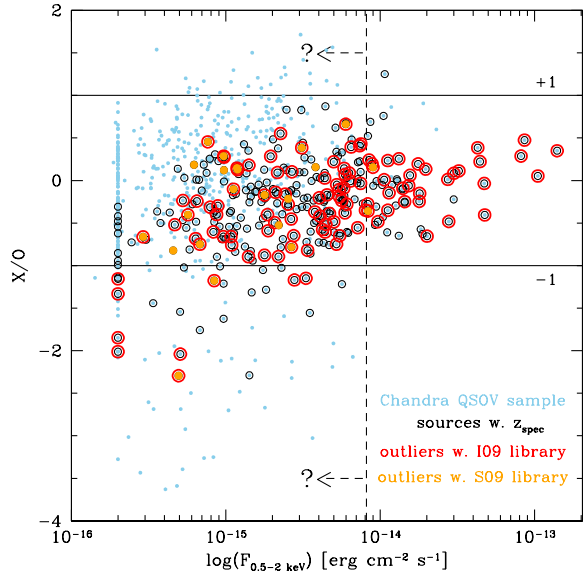} 
 
\caption{ \small As in Figure\,\ref{fig:extnv} but for C-COSMOS {\it QSOV} sample. Clearly the templates for normal galaxies are unsuitable for this sample.}
\label{fig:qsov}
\end{center}
\end{figure}

In an attempt to understand the origin of the systematic errors, as  in the previous section, we plot in Figure\,\ref{fig:qsov} the outliers obtained using the S09 library (yellow filled circles) as a function of optical and X-ray brightness. For  the sake of completeness, we also plot the outliers that are obtained by using the library of normal galaxies from I09 (red circles), imposing this time the luminosity prior -20$<\rm M_B<$-30. We demonstrated in S09 that this library is unsuitable for the XMM {\it QSOV} sample, but in Table\,\ref{tab:qsov} and Figure\,\ref{fig:qsov}  this can be seen more clearly. For the {\it QSOV} sample, the library of AGN templates helps to measure more accurate photo-z than the library of normal galaxies at any X-ray flux and any optical magnitude.

Thus, the main limitation  to the accuracy of the photo-z and the fraction of outliers appears to be related to the optical faintness of the sources.  As already pointed out by other authors \cite[e.g.,][]{Cardamone:2010lr, Barro:2011rt}, at fainter magnitudes the  spectral energy distribution is  less tightly constrained, and only by upper limits in some bands, or has large statistical uncertainties associated with the photometry. Thus, the 1$\sigma$ error associated  with $z_{\rm phot}$ steadily increases with the {\it i}$^*$ band magnitude for the COSMOS multi-wavelength data-set (see S09, I09).
Only deeper photometry in NIR bands (where the 4000 \AA\, break falls 2$<$z$<$2.5) will allow us  to improve the accuracy of the photo-z for the faint sources in the C-COSMOS and XMM-COSMOS {\it QSOV} samples. Thus, an opportunity to improve the results for at least a fraction of the sources will be given with the photometry from ULTRAVISTA survey\footnote{\url{http://www.strw.leidenuniv.nl/$\sim$ultravista/}} and the future observations taken with HST/WFC3 by the CANDELS survey \citep{Grogin:2011fr, Koekemoer:2011gf}.
  
 
\section{Results}
\label{sec:results}

\begin{figure}[ht]
\begin{center}

\includegraphics[scale=1]{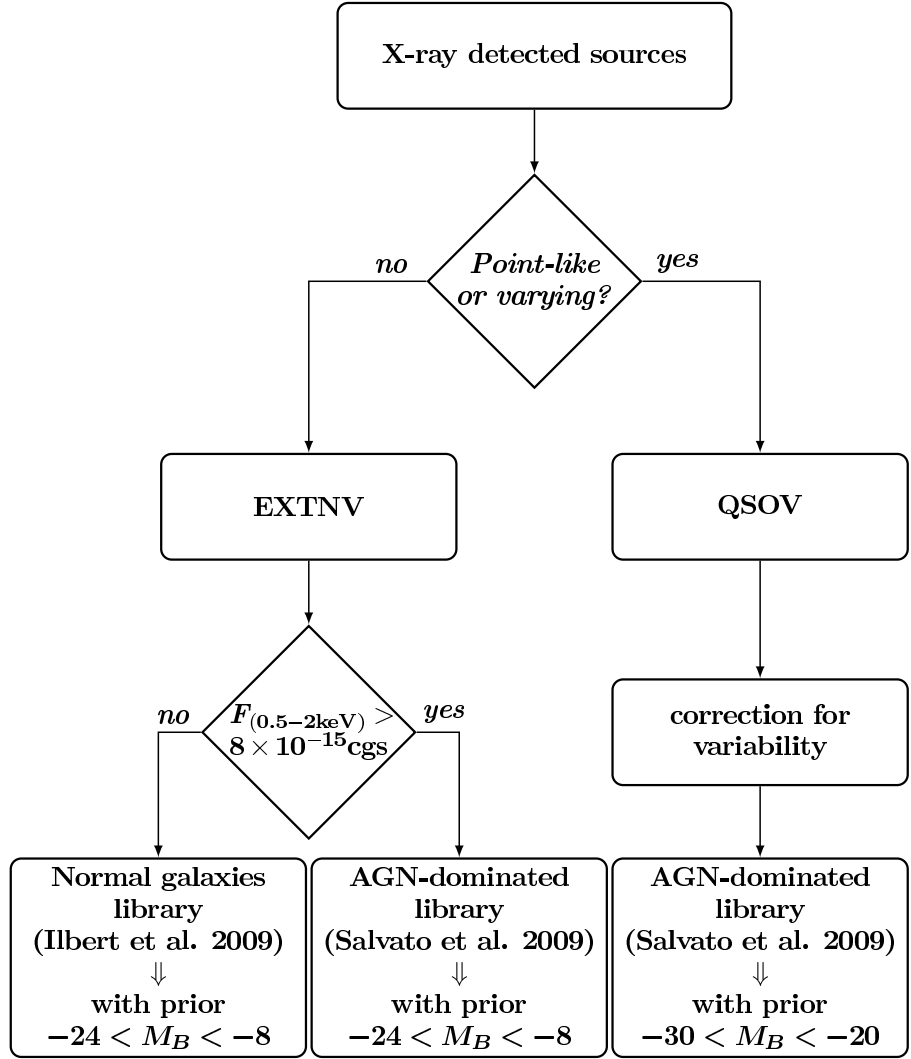} 
\caption{ \small Flow-chart of the procedure adopted to compute photo-z for X-ray detected sources.}

\label{fig:flowchart}
\end{center}
\end{figure}

\begin{figure*}
\begin{center}
\includegraphics[scale=0.4]{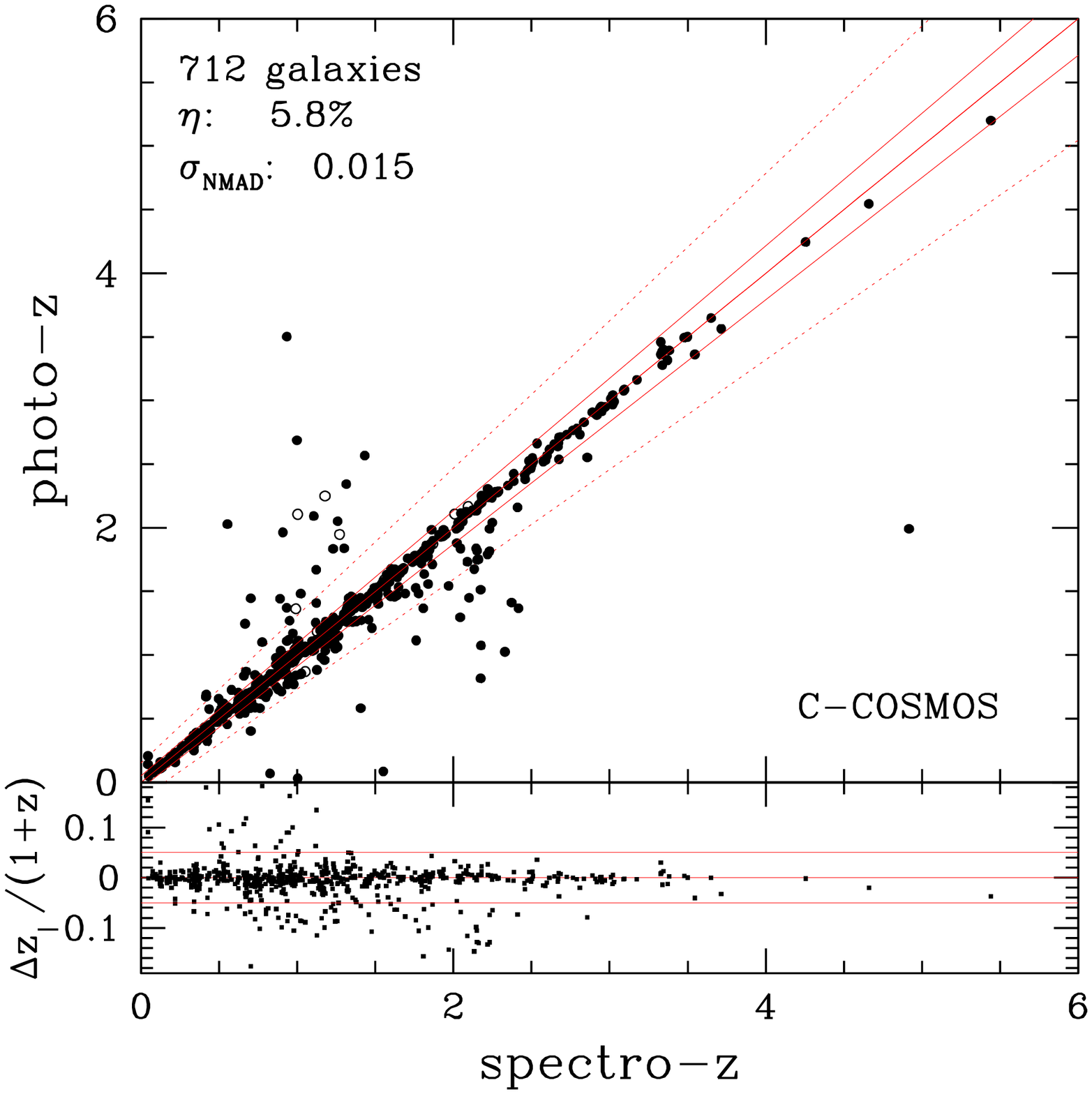}  
\includegraphics[scale=0.4]{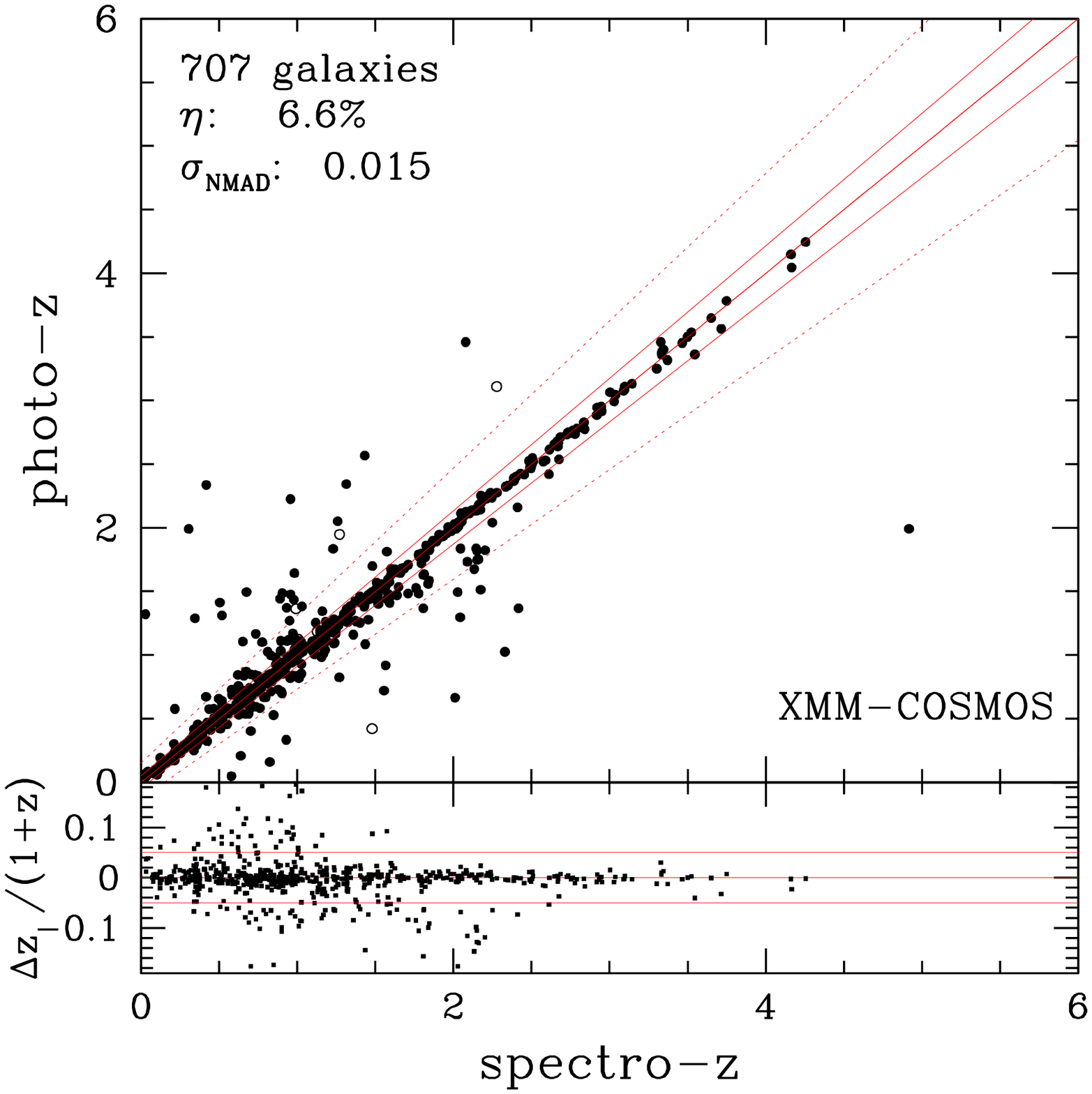}

\caption{ \small Final photometric vs spectroscopic redshifts for the entire  C-COSMOS (Left) and XMM-COSMOS (Right) samples. }
\label{fig:zphot_final}
\end{center}
\end{figure*}

In summary, using the procedure described in the paper and illustrated in the flow-chart of Figure\,\ref{fig:flowchart} we obtained high quality photo-z for C-COSMOS. In addition,  we recomputed the photo-z  of XMM-COSMOS  sources for which the $H$-band photometry is now available. 
For both samples (see Figure\,\ref{fig:zphot_final}), we obtained an accuracy of $\sigma_{NMAD}$=0.015 and a similar fraction of outliers $\eta\sim$6\%.
The addition of $H$-band photometry and  our revised strategy for treating the extended, optically non-varying, faint X-ray sources in the {\it EXTNV} XMM-COSMOS sample, resulted in a change in photo-z of $\Delta z>$ 0.2 for 248 sources ($\sim$ 15\% of the total XMM-COSMOS sample). This improved accuracy with respect to the old version of the photo-z catalog \citep{Salvato:2009zw} is summarized in Table~\ref{tab:old-new-xmm}. It is reassuring that the introduction of the $H$-band photometry does not  affect the accuracy of the {\it QSOV} sample,  illustrating the reliability of our photo-z in the field.

The final photo-z catalogs for the C-COSMOS and XMM-COSMOS surveys are  available\footnote{ \url{http://www.ipp.mpg.de/$\sim$msalv/PHOTOZ\_XCOSMOS/}} in ASCII format, together with morphological and variability analysis.
 Excerpt of the catalogs are provided in Table \ref {tab:summary_example}.
 

In both catalogs, we  flagged as {\it stars} those sources that are point-like and have $1.5 \times \chi^2_{\rm star} < \chi^2_{\rm agn/gal}$, where $\chi^2_{\rm star}$ and $\chi^2_{\rm agn/gal}$ are the reduced $\chi^2$ for the  best-fit solutions obtained with stellar and AGN or galaxy libraries.
For C-COSMOS(XMM-COSMOS), we found 33(53) candidate stars, 18(32) of which are already spectroscopically confirmed. The criterion  fails to identify 5(3)  sources that are known to be stars via spectroscopy.
A more relaxed criterion,  such as the one used in I09 and S09 (point-like and  $\chi^2_{\rm star} < \chi^2_{\rm agn/gal}$) would allow the identification of all the spectroscopic stars, but would also misclassify as stars objects that are spectroscopically confirmed galaxies.
\\

\begin{figure}
\begin{center}
\includegraphics[scale=0.45]{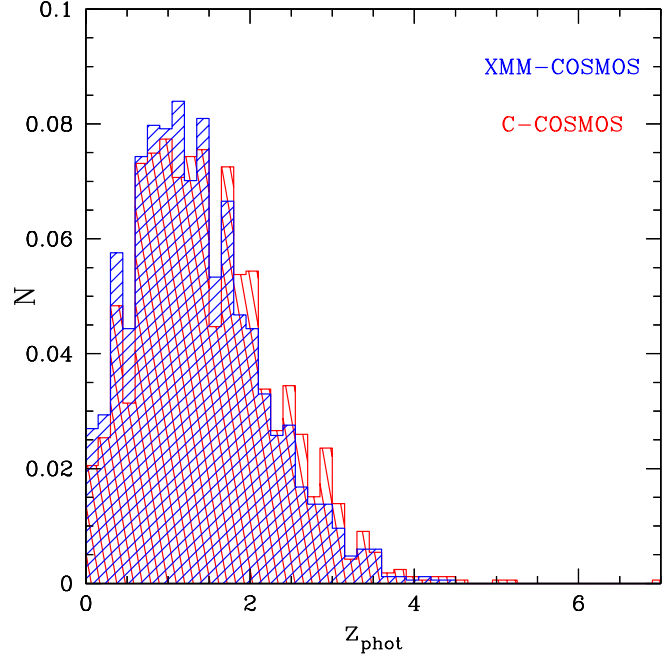}  

\caption{ \small  Photo-z distribution for C-COSMOS (red) and XMM-COSMOS (blue), normalized to the respective total number of sources.}
\label{fig:histo_z_all.eps}
\end{center}
\end{figure}

The redshift distribution for the galaxies in the two samples (red: C-COSMOS; blue: XMM-COSMOS) is shown in Figure\,\ref{fig:histo_z_all.eps}, where the histograms are normalized to the respective total number of sources. As expected, the deeper X-ray observations  of C-COSMOS allowed  us to detect sources at higher redshift (from z$\sim$1.8) than XMM-COSMOS.  

\begin{figure}
\begin{center}
\includegraphics[scale=0.4]{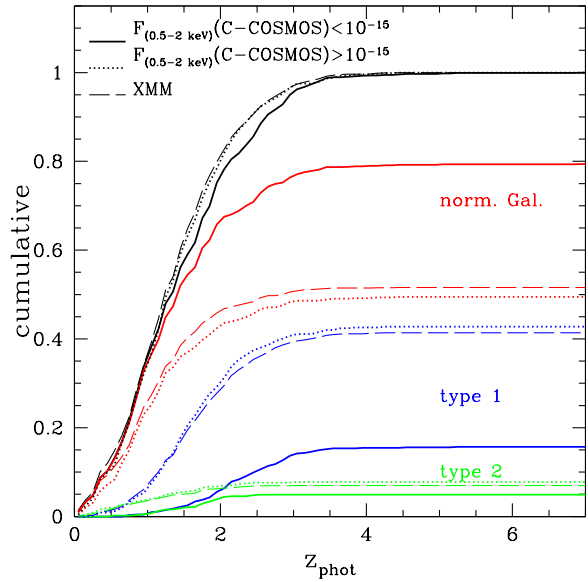} 
\caption{ \small Normalized, cumulative redshift distribution for all XMM-COSMOS sources (black dashed line) and C-COSMOS sources with soft X-ray flux below (black solid line) or above (black dotted line) the detection limit of XMM-COSMOS (F$_{\rm 0.5-2 keV} =10^{-15} {\rm erg~cm}^{-2} {\rm s}^{-1}$).
Red lines indicate the sources  more accurately described by templates of normal galaxies, while blue and green lines indicate sources better fit by type 1 AGN and type 2 AGN, respectively.}
\label{fig:ztemplates}
\end{center}
\end{figure}

To assess any other differences in the populations of the two surveys, we considered the C-COSMOS sources that are respectively below (C-COSMOS faint) and above (C-COSMOS bright) the flux limit of XMM-COSMOS (F$_{0.5-2 keV} =10^{-15} {\rm erg~cm}^{-2} {\rm s}^{-1}$ ). The comparison between the two sub-samples and the XMM-COSMOS survey is shown in Figure\,\ref{fig:ztemplates}, where the black solid line represents the redshift distribution of the C-COSMOS faint sources and the thick, dashed and thin, dotted black lines represent C-COSMOS bright and XMM-COSMOS sources, respectively. 
While C-COSMOS bright and XMM-COSMOS do not differ, the Kolmogorov-Smirnov (KS) test  suggests that the population of sources in C-COSMOS bright and faint are not extracted from the same parent population (P$_{\rm H_0}\sim$ 0.006\% ), as already appeared to be clear from the previous Figure\,\ref{fig:histo_z_all.eps}.

In our additional analysis, we divided the samples according to the  best-fit SED template. Red lines trace the cumulative distributions of the sources fitted by normal  galaxy templates, while green and blue lines indicate  sources that can be most accurately described by type 2 AGN and type 1 AGN  templates, as defined in S09.
The KS test gives a probability of P$_{\rm H_0}\sim 0.001$, P$_{\rm H_0}\sim 0.014$, and P$_{\rm H_0}\sim 0.011$ that the  three populations (galaxies, type 1, and type 2 AGN) are  drawn from the same population for C-COSMOS and XMM-COSMOS.
This is somehow implicit in the procedure used  to estimate the photo-z, as we change  our library in accordance with the X-ray flux for the {\it EXTNV} samples and C-COSMOS being deeper than XMM. We fitted respectively 90\% and 95\% of the {\it EXTNV} C- and XMM-COSMOS  sources using the library of normal galaxies.   

However,  we note that even if these sources are  more accurately described by normal galaxy templates, they are  not normal galaxies.  We can more accurately describe 994(935) of the 1098(1045) C-COSMOS(XMM-COSMOS) sources ($\sim$91\%) using a normal galaxy template but  these sources have X-ray luminosities above 10$^{42} $ erg/s and thus can be  assumed to be powered by an active nucleus.  
 
 
\subsection{The highest redshift X-ray selected sources?}
\label{subsec:highz}

 By combining  their spectroscopic and photometric  data, \cite{Civano:2011qy} presented the logN-logS and space density of C-COSMOS high-z sources (z$>$3), and we refer to this paper for a detailed discussion of the high-z X-ray source population. Here we present the photometric properties of the highest-z X-ray selected candidate AGN and investigate the effects of different assumptions  about the SED templates and luminosity priors on the photo-z  estimation, and its stability and reliability.

High-redshift AGN provide key observational constraints of the theoretical models of galaxy and SMBH formation and evolution. Most models can well describe the high-luminosity regime up to z $\sim$2-3 \citep[][and references therein]{Hopkins:2008qy,Menci:2008uq}. However, the shortage of observational data for both high-redshift and low-luminosity AGN populations has  restricted the progress of the modeling.  Since these predictions are generally applied to determine the key physical parameters such as the QSO duty cycle, the black hole seed mass function, and the accretion rates, reliable observations of the QSO luminosity function and its evolution at high redshift are required \citep[Civano et al 2011,][]{Brusa:2010lr, Aird:2010fj, Fontanot:2007lr}.

\begin{figure*}
\begin{center}
\hspace*{-0.5truecm}\includegraphics[scale=0.2]{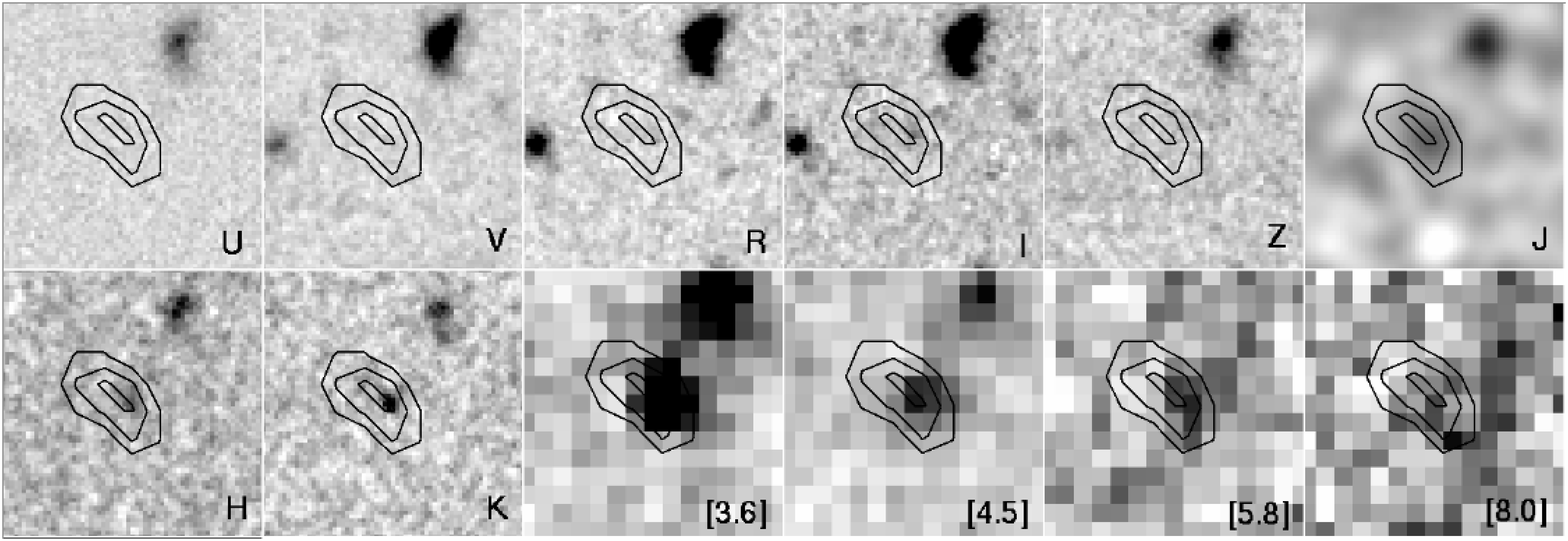}
\includegraphics[scale=0.3]{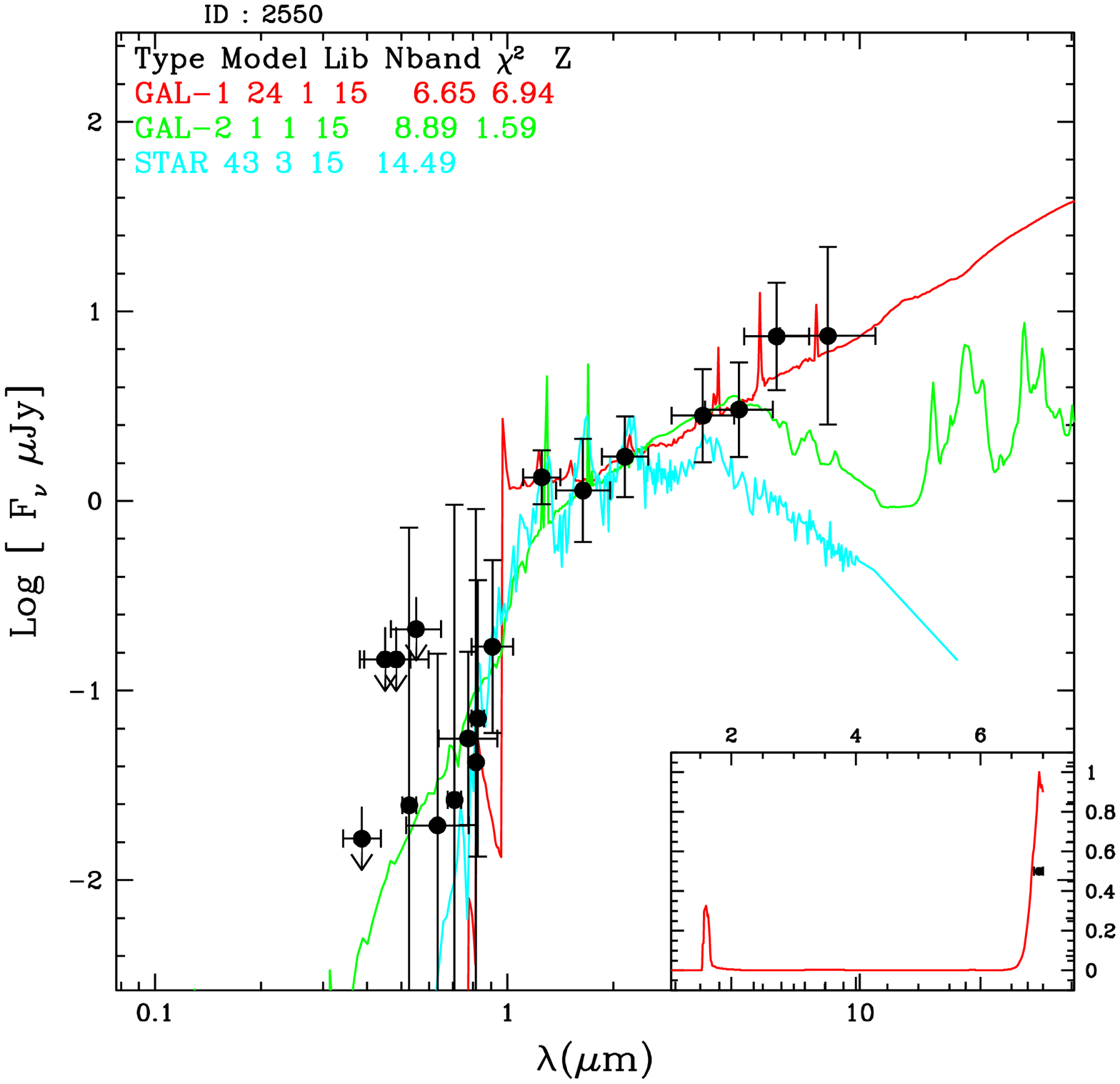}
\caption{ \small {\bf Left:} From left-top to right-bottom, stamp images (10"x10") in U, V, R, I, Z, J, H, K, and the 4 IRAC channels (3.6, 4.5, 5.8, 8 $\mu {\rm m}$) for CID-2550. Black contours indicate the X-ray detection. The source is  clearly visible in the bands redder than 9000$\AA$. {\bf Right:} Spectral energy distribution of source CID-2550. Depending on the adopted luminosity priors,  one or two photo-z solutions are found, although the low redshift solution has always very low PDFz. }
\label{fig:highZCandidate}
\end{center}
\end{figure*}

 In addition to the 19 sources that, as discussed in \S~\ref{sec:data}, are potentially at z$_{\rm phot}>$4, the C-COSMOS sample contains a single source for which the  most likely photo-z solution is at z$>$6. In contrast to the typical results for high-z candidate sources, the redshift probability distribution function (PDFz) is  both peaked and narrow.
The counterpart  to the source CID-2550 is detected long-wards of 9000\,\AA ($z_{\rm AB}=25.4$ mag; $J_{\rm AB}=23.6$ mag; $H_{\rm AB}=23.8$ mag; 
$K_{\rm AB}=23.0$ mag; $[3.6\mu {\rm m}]_{\rm AB}=22.8$ mag; $[4.5 \mu {\rm m}]_{\rm AB}=22.7$ mag; $[5.8\mu {\rm m}]_{\rm AB}=21.7$ mag; $[8\mu {\rm m}]_{\rm AB}=21.7$ mag; see left panel of Figure~\ref{fig:highZCandidate}) and is also marginally seen in the deep Subaru $i^+$-band observations ($\approx 26.6$\,mag).

Photo-z are usually very sensitive to luminosity priors, in particular when the available photometry has large uncertainties and/or the number of the photometric points are insufficiently large to reliably determine a photo-z. This is case for CID-2550 where in the optical bands we have either an upper limit or errors larger than 1 magnitude. Imposing a lower limit  to the absolute magnitude of M$_B=-20.5$ results in a unique (PDFz=98\,\%) solution at $z_{\rm phot}=6.84$ with a best-fit SED solution being obtained using an AGN+ULIRG hybrid (QSO1+IRAS22491, see S09).
Without any luminosity prior, the best-fit photo-z solution becomes $z_{\rm phot}=6.94$ (PDFz=85\,\%) with a second, less probable solution at $z_{\rm phot}=1.59$ (see Figure\,\ref{fig:highZCandidate}, where, in cyan, we  also plot the best fit obtained with a library of stars). The template that most accurately describes the data remain the same, while for the low redshift solution a dusty blue SB template from \cite{Bruzual:2003uq} is preferred.
 
The high redshift solution suggested by the PDFz is also supported by the very small number of outliers that we obtain at high redshift (only 1 out of 53 sources at $2.5<z_{\rm spec}<5.4$, with $\sigma_{NMAD}=0.009$). In addition, a solution at z=6.84 would explain the marginal detection in the deep Subaru $i^+$-band  as emission from ${\rm Ly}_{\beta}$  caused by an incomplete Gunn-Peterson trough  \citep{Becker:2001lr,Fan:2006fk}.
The  source is also comparable to the Extreme X-ray/Optical ratios  sources (EXOs) first defined by \cite{Koekemoer:2004kx} which are selected as optical dropouts with X-ray emission, although the improved multi-wavelength data available  here for CID-2550 provide a stronger photo-z solution. 

At $z\sim6.84$, the 0.5-7\,keV X-ray luminosity for CID-2550 would be log(L$_X$)$=44.67$\,erg s$^{-1}$, while the absolute B-band magnitude would be $M_B=-24.6$, i.e.  a significantly high QSO luminosity.
Assuming a) that the quasar emits at the Eddington luminosity, b) an X-ray bolometric correction in the range 10-100, and c) that neither lensing  nor beaming  significantly magnify the observed flux, we estimate a central black hole  mass in the range  $\approx 4\times10^{7...8}\,M_{\odot}$.  This mass estimate is lower than the average mass derived for the sample of bright optically-selected $z>6$ quasars from SDSS \citep{Fan:2001ys,Willott:2003qy}, suggesting that X-ray selection might detect less extreme objects, or   objects in a different, possibly obscured (as suggested by the best fit hybrid galaxy template), phase of rapid growth.
   

\section{Discussion}
\label{sec:discussion}

\subsection{Importance of spectroscopic sample}
Most galaxy and AGN (co)evolutionary studies  depend on photo-z estimates. Spectroscopy is extremely challenging, in particular, at high redshift, where photo-z then play a fundamental role. The photo-z accuracy is usually estimated by  comparing with a small spectroscopic sample of bright and/or nearby objects.  Both the telescope diameter and  the wavelength coverage of the spectrographs dictate the parameter range here.

For bright and nearby sources, the photometric coverage is comprehensive and the data accurate, making the computation of reliable photo-z relatively easy.
In contrast, with increasing faintness of the --possibly at high redshift-- sources , the spectral energy distributions become less  clearly defined \citep[e.g.,][]{Hildebrandt:2008uq}.  
 Fewer reliable source detections,  larger statistical uncertainties associated with photometry, and an increasing number of upper limits, lead to poorly constrained SEDs (I09, S09).
While this affects normal galaxies and AGN  in similar ways, the situation for the latter is complicated by the uncertainty  in the relative  contributions of  the nuclear and host emission components.

These uncertainties were considered in \S\,\ref{subsec:asXMM}, where we presented the application of our photo-z procedure to the XMM-COSMOS survey  and  the deeper C-COSMOS sample, and illustrated its limitations in correctly reproducing the properties of the faint end of the flux distribution.

For XMM-COSMOS, a large training spectroscopic sample allowed us to  characterize the bright sources extremely well.
Thus, when the same procedure was applied to the {\it Chandra} sources with similarly bright optical counterparts ({\it i}$^*_{AB}<$22.5),
it  provided a comparable accuracy and no further tuning of the library or the priors was required.

For the faint counterparts in XMM-COSMOS, no statistically meaningful spectroscopic sample was available, thus no tuning for these sources was performed. The good agreement of the photo-z for the few XMM faint sources with spectroscopic redshifts, suggested that the setup for the bright population could be extended to the entire X-ray sample.
However, the significant increase in the spectroscopic sample with {\it i}$^*_{AB}>$22.5 for C-COSMOS indicated that the results for XMM-COSMOS in this range were likely the outcome of small  number statistics and that a  more careful study of the faint sub-sample was needed.

This demonstrated again the importance of the choice of the training sample for the quality of the photo-z. The accuracy and number of outliers calculated for a set of sources with spectroscopy can be used as a quality indicator for photo-z only if the sample without spectroscopy covers the same parameter space as the training sample. For a population dominated by sources fainter than the spectroscopic training sample, the quality of the photo-z is often overestimated.

\subsection{Application to other X-ray surveys}
\label{subsec:generalization}

The strength of any photo-z estimation method is reflected most by how generally it can be applied. In \S\,\ref{sec:results}, we illustrated how the procedure developed here for C-COSMOS led to  an improvement in the photo-z for XMM-COSMOS (see Figure\,\ref{fig:zphot_final} and Table\,\ref{tab:old-new-xmm}).
For a similar test, we applied the method to the sources detected by XMM in another deep field, the Lockman Hole  (Fotopoulou et al. 2011, submitted). The photometric coverage of the Lockman Hole has been extended to 22 broad bands from UV to mid-infrared \citep{Rovilos:2009fj, Rovilos:2011yq}, together with deep HST/ACS imaging. With these data, we have been able to reach an accuracy of $\sigma_{NMAD}=0.07$ and a fraction of outliers $\eta$=12.5\%. The two values are comparable to the results of C-COSMOS, if the same photometric bands and depths as used for the Lockman Hole are used and no variability correction  is  applied.

This suggests that our procedure is robust and that its level of success is now dictated  only by the available filters and depths. The procedure can be straightforwardly applied to the large number of deep multi-wavelength pencil-beam X-ray surveys  such as, for example, AEGIS-X \citep{Laird:2009fj} and CDFS  \citep{Luo:2010qp} or E-CDFS \citep{Cardamone:2010lr}.
The study of the AEGIS-X field, which covers 0.67 square degrees, would benefit greatly from our procedure as the field is a) wide enough to include some bright AGN (needing AGN templates) and b) deep enough (F$_{\rm 0.5-2 keV} =5.3 \times 10^{-17} {\rm erg~cm}^{-2} {\rm s}^{-1}$) to include sources that are AGN but for which the SED is more closely fit by normal galaxy templates.
In addition, an accurate and merged photometric AEGIS-X catalog is now available \citep{Barro:2011rt}.
 
E-CDFS (0.25 square degrees) is probably the most deeply observed portion of sky in terms of both imaging and spectroscopy. This  has allowed a reconstruction of the SEDs of the sources and a knowledge of the flux-redshift parameter space also at faint magnitudes.  For the X-ray detected sources, reliable photo-z has become available \citep{Luo:2010qp,Cardamone:2010lr}. By applying different methods and  using partially different datasets   (such as the additional photometry from 18 deep intermediate-band filters in Cardamone et al.), both groups obtain an accuracy of $\sigma<0.01$. However, for 75 of the 169 sources without spectroscopy (i.e. 44\%) that they have in common, the photo-z values differ by more than 0.2.
 Once again, it is clear that a good match between photometric and spectroscopic redshifts is not synonymous of univocal results.
   
Crucial information about the X-ray source population in the Universe will also be provided by wide-field and all-sky missions, such as the {\it eROSITA}  mission \citep{Cappelluti:2011qy},  which is planned  for launch in 2013 and is expected to detect several millions of AGN brighter than F$_{\rm 0.5-2 keV} =10^{-14} {\rm erg~cm}^{-2} {\rm s}^{-1}$ in the all-sky. This flux limit is about a factor of 50(10) brighter than the C-COSMOS (XMM-COSMOS) limit and the contamination by X-ray emitting normal galaxies is likely negligible. Thus, the color-redshift degeneracy could in principle almost be eradicated by using the  AGN-dominated S09 library for both the {\it QSOV} and the {\it EXTNV} samples.

\begin{figure*}
\begin{center}
\includegraphics[scale=0.4]{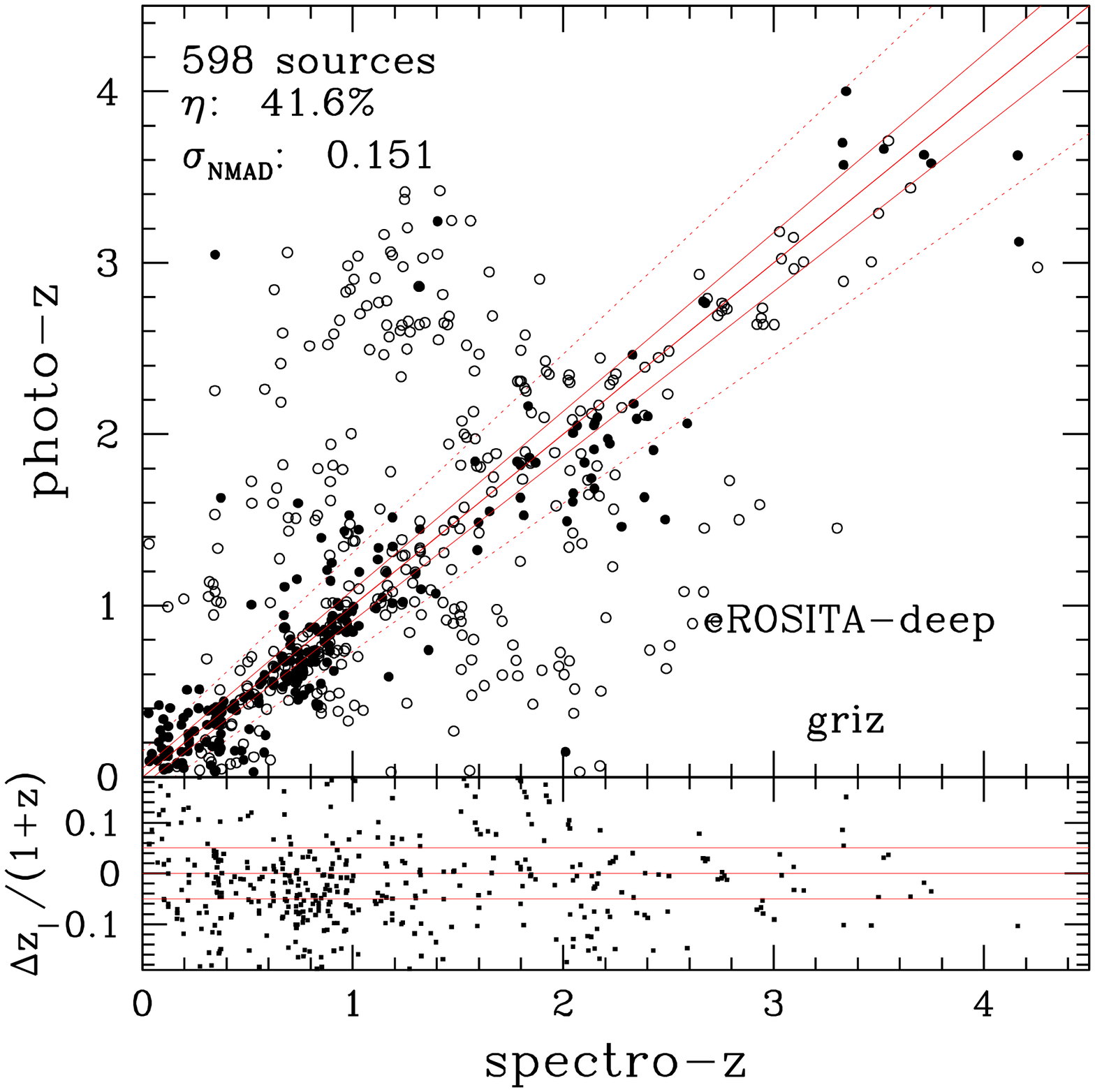} 
\includegraphics[scale=0.4]{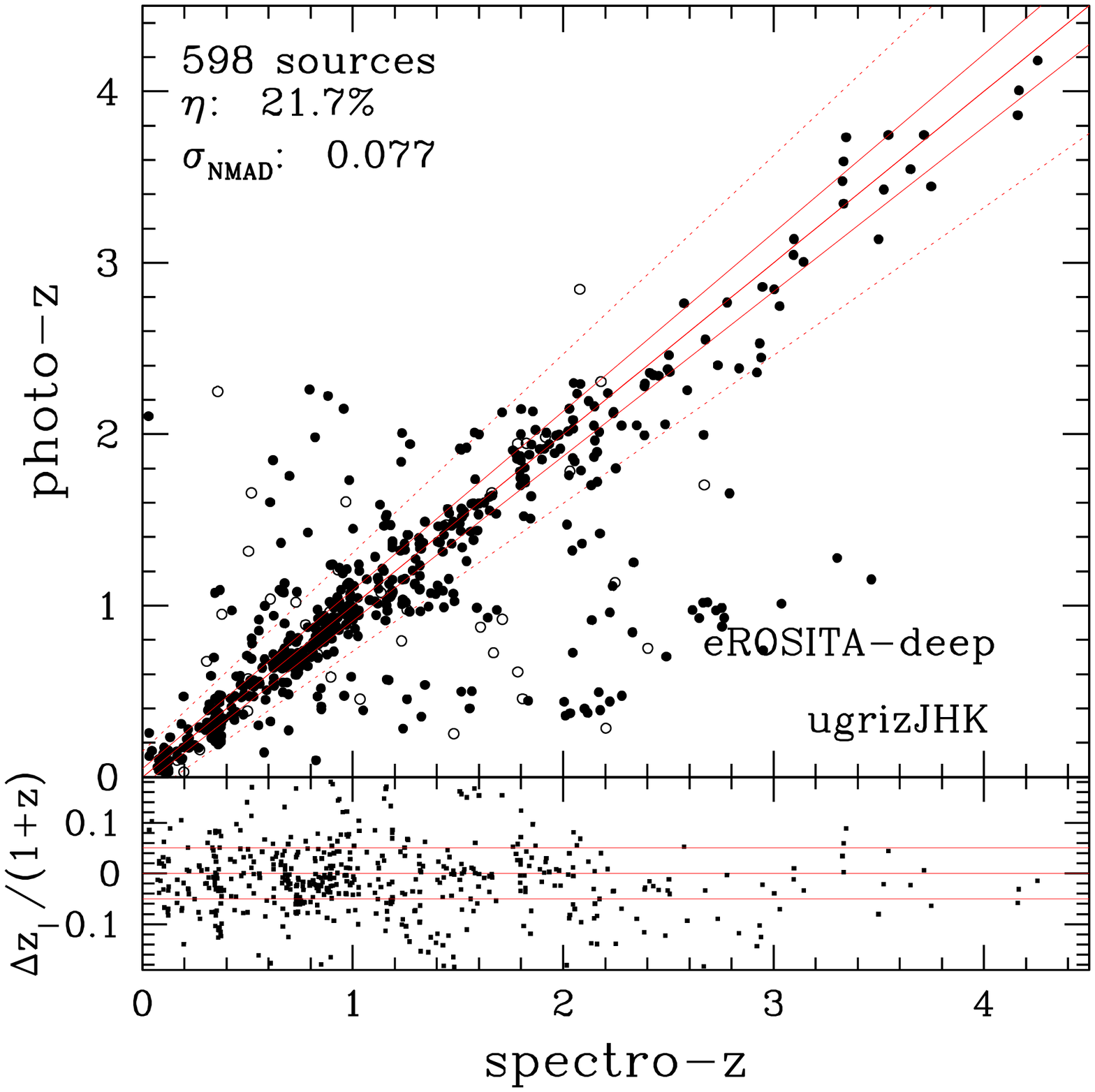} 
\caption{ \small Photometric vs spectroscopic redshifts for XMM-COSMOS sources at the X-ray depth of eROSITA Deep  (F$_{\rm 0.5-2 keV} =1 \times 10^{-15} {\rm erg~cm}^{-2} \rm s^{-1}$), using  {\it ``griz''} broad-band photometry (Left panel ) and ``ugrizJHK'' (Right panel). The high dispersion and fraction of outliers would rend the photo-z computed with four bands, in the traditional way, unusable.  Only the addition of ``u''  and ``JHK'' would allow reasonable results. This option should be considered at least for the deep part of the eROSITA survey. }
\label{fig:eROSITA_ugriz}
\end{center}
\end{figure*}

However, the deep optical all-sky bands useful for the identification of the {\it eROSITA} sources will likely be limited to 4-5 broad bands from Pan-STARSS \citep{Burgett:2009fr}, LSST \citep{Ivezic:2006mz}, Skymapper \citep{Tisserand:2008fj}, and DES \citep[][]{DePoy:2008fk,Mohr:2008qy}.  While such a SDSS-like filter set can help to provide reliable photo-z for normal galaxies up to z$\sim$1 \citep{Oyaizu:2008lr},  it is insufficient for AGN. In the left panel of Figure\,\ref{fig:eROSITA_ugriz}, we compare the photometric and spectroscopic redshifts  of mock {\it eROSITA} sources.  
Here, we used the XMM-COSMOS sample cut at the X-ray flux above F$_{\rm 0.5-2 keV} =10^{-15} {\rm erg~cm}^{-2} {\rm s}^{-1}$ ({\it eROSITA} depth planned for the 2 $\times$ 100 square degrees deep areas) and computed the photo-z using only {\it griz} photometry, as  will be available from a very deep (i=26 mag) Pan-STARRS filter set  after a correction for variability. As expected, the fraction of outliers is large ($\eta$=41.5\%) and the accuracy is well below what one would wish to achieve($\sigma_{NAMD}\sim 0.150$ for sources brighter than  F$_{\rm 0.5-2 keV} =10^{-15} {\rm erg~cm}^{-2} {\rm s}^{-1}$; $\sigma_{NAMD}\sim 0.24$ for sources brighter than F$_{\rm 0.5-2 keV} =10^{-14} {\rm erg~cm}^{-2} {\rm s}^{-1}$ ). Only with the addition of the ``u'' and ``JHK'' (right panel of Figure\,\ref{fig:eROSITA_ugriz})  photometry will we be able to reach an accuracy that would allow us to use the measured photo-z for scientific studies. Without variability correction the fraction of outliers would increase by additional 10\%.


This  clearly demonstrates the importance of multi-epoch observations and well-sampled SEDs \citep[see also][for simulations]{Benitez:2009lr}.  The availability  of only broad-band photometry  will greatly limit the possibility of using  SED- fitting for computing photo-z for AGN and  new methods should be,  such as the inclusion of additional priors  as the redshift or flux-redshift distributions (\cite[e.g.][]{Benitez:2000it} and \cite{Bovy:2011qy}, respectively. Only in this way will future X-ray surveys  be able to maximize  the insight they achieve in understanding AGN/galaxy (co)evolution.

\section{Conclusions}
\label{sec:conclusions}
 
 It is generally believed that AGNs are playing a major role, although still to be fully understood, for galaxy formation and evolution.
However, AGNs are rare compared to galaxies. Thus, assembling large AGN redshift samples is a real challenge, and requires much more telescope time than acquiring photometric data. As a consequence, the main motivation for our work is the development of a better wayÊ to measure accurate photo-z forÊ AGN-dominated galaxies using large photometric surveys. 
 
In this paper, we have presented,  and tested  thoroughly our methodology to derive photometric redshifts for X-ray sources.  Our robust tuning of the photo-z technique for AGN has been made possible thanks to a) the sizable training spectroscopic sample spanning Êa large range in redshift, luminosity, and morphology of sources, b) the multi-wavelength coverage,Ê and c) the correction for variability effects. 

We presented the photo-z measurements of 1692 {\it Chandra} detected sources andÊ 1683 XMM detected sources in the COSMOS field (869 sources are common to both surveys). While the former surveyÊ covers the central central 0.9\,deg$^2$ atÊ a depth of $F_{\rm (0.5-2\,keV)}$= $1.9\times10^{-16}$ erg cm$^{-2}$ s$^{-1}$, the latter is a factor of 3-4 shallower butÊ covers the entire 2 square degreesÊ of the COSMOS field. For both samples, we have achieved an accuracy of $\sigma_{NMAD}$=0.015 and a fraction of outliers $\eta \sim$ 6\%.  In comparison with our previous analysis on the XMM-COSMOS sample (Salvato et al. 2009), we have shown that better results are obtained for faint, extended sources, that do not display optical variability, when a library of normal galaxies is used to fit their SED.

We have argued that the photo-z procedure adopted forÊ X-ray sources in COSMOS can be applied to other X-ray surveys and will be a major asset for the scientific exploitation of any future large X-ray programs. The achievable accuracy is now limited only byÊ both the depth of the photometric data and the number of the photometric bands available. For this reason, we propose that wide/all-sky X-ray surveys should invest substantially in multi-wavelength follow-up observationsÊ to enable researchers to fully exploit the potential of these surveys in studying AGN evolution.

\acknowledgments

We gratefully acknowledge the contributions of the entire COSMOS collaboration consisting of more than 100 scientists. More information  about the COSMOS survey is available at http://www.astro.caltech.edu/$\sim$cosmos. We also acknowledge the use of STILTS and TOPCAT tools \citep{Taylor:2005rt}. We acknowledge the anonymous
referee for helpful comments that improved the paper. MS and GH acknowledge support by the German Deutsche Forschungsgemeinschaft, DFG Leibniz Prize (FKZ HA 1850/28-1). FC was supported in part by NASA Chandra grant
number GO7-8136A, the Blancheflor Boncompagni 
Ludovisi foundation and the Smithsonian Scholarly Studies.AC,CV,NC,FF acknowledge financial contribution from the agreement ASI-INAF I/009/10/0.

{\it Facilities:} \facility{KECK}, \facility{HST}, \facility{VLT},\facility{CXO},\facility{SDSS},\facility{XMM},\facility{Subaru}.






\bibliographystyle{apj}

\begin{thebibliography}{}


\bibitem[{{Aird} {et~al.}(2010){Aird}, {Nandra}, {Laird}, {Georgakakis},
  {Ashby}, {Barmby}, {Coil}, {Huang}, {Koekemoer}, {Steidel}, \&
  {Willmer}}]{Aird:2010fj}
{Aird}, J., {Nandra}, K., {Laird}, E.~S., et al., 2010, \mnras, 401, 2531

\bibitem[{{Arnouts} {et~al.}(1999)}]{Arnouts:1999lr}
{Arnouts}, S., {Cristiani}, S., {Moscardini}, L. et al., 1999, \mnras, 310, 540

\bibitem[{{Barro} {et~al.}(2011){Barro}, {P{\'e}rez-Gonz{\'a}lez}, {Gallego},
  {Ashby}, {Kajisawa}, {Miyazaki}, {Villar}, {Yamada}, \&
  {Zamorano}}]{Barro:2011rt}
{Barro}, G., {P{\'e}rez-Gonz{\'a}lez}, P.~G., {Gallego}, J., et al., 2011, \apjs, 193, 13

\bibitem[{{Becker} {et~al.}(2001){Becker}, {Fan}, {White}, {Strauss},
  {Narayanan}, {Lupton}, {Gunn}, {Annis}, {Bahcall}, {Brinkmann}, {Connolly},
  {Csabai}, {Czarapata}, {Doi}, {Heckman}, {Hennessy}, {Ivezi{\'c}}, {Knapp},
  {Lamb}, {McKay}, {Munn}, {Nash}, {Nichol}, {Pier}, {Richards}, {Schneider},
  {Stoughton}, {Szalay}, {Thakar}, \& {York}}]{Becker:2001lr}
{Becker}, R.~H., {Fan}, X., {White}, R.~L., et al., 2001, \aj, 122, 2850

\bibitem[{{Bender}}{et~al.} (2001){Bender}, {Appenzeller}, {B\"ohm}, {Drory}, {Fricke}, {Gabasch}, {Heidt}, {Hopp}, {J\"ager}, {K\"ummel}, {Mehlert}, {M\"ollenhoff}, {Moorwood}, {Nicklas}, {Noll}, {Saglia}, {Seifert}, {Seitz}, {Stahl}, {Sutorius}, {Szeifert}, {Wagner}, { Ziegler}]{Bender:2001fk}
{Bender}, R., {Appenzeller}, I., {B\"ohm}, A.~et al. 2001, Proceedings of the ESO Workshop Held at Garching, Germany, 9-12 October 2000, ESO ASTROPHYSICS SYMPOSIA. ISBN 3-540-42799-6. Edited by S. Cristiani, A. Renzini, and R.E. Williams. Springer-Verlag, 2001, p. 96


\bibitem[{{Ben{\'{\i}}tez}(2000)}]{Benitez:2000it}
{Ben{\'{\i}}tez}, N. 2000, \apj, 536, 571

\bibitem[{{Ben{\'{\i}}tez} {et~al.}(2009){Ben{\'{\i}}tez}, {Moles}, {Aguerri},
  {Alfaro}, {Broadhurst}, {Cabrera-Ca{\~n}o}, {Castander}, {Cepa},
  {Cervi{\~n}o}, {Crist{\'o}bal-Hornillos}, {Fern{\'a}ndez-Soto}, {Gonz{\'a}lez
  Delgado}, {Infante}, {M{\'a}rquez}, {Mart{\'{\i}}nez}, {Masegosa}, {Del
  Olmo}, {Perea}, {Prada}, {Quintana}, \& {S{\'a}nchez}}]{Benitez:2009lr}
{Ben{\'{\i}}tez}, N., {Moles}, M., {Aguerri}, J.~A.~L., et al., 2009, \apjl, 692, L5

\bibitem[{{Bertin} \& {Arnouts}(1996)}]{Bertin:1996qy}
{Bertin}, E. and {Arnouts}, S., 1996, \aaps, 117, 393

\bibitem[{{Bovy} {et~al.}(2011){Bovy}, {Myers}, {Hennawi}, {Hogg}, {McMahon},
  {Schiminovich}, {Sheldon}, {Brinkmann}, {Schneider}, \&
  {Weaver}}]{Bovy:2011qy}
{Bovy}, J., {Myers}, A.~D., {Hennawi}, J.~F., {Hogg}, D.~W., et al., 2011, ArXiv e-print


\bibitem[{{Brusa} {et~al.}(2010){Brusa}, {Civano}, {Comastri}, {Miyaji},
  {Salvato}, {Zamorani}, {Cappelluti}, {Fiore}, {Hasinger}, {Mainieri},
  {Merloni}, {Bongiorno}, {Capak}, {Elvis}, {Gilli}, {Hao}, {Jahnke},
  {Koekemoer}, {Ilbert}, {Le Floc'h}, {Lusso}, {Mignoli}, {Schinnerer},
  {Silverman}, {Treister}, {Trump}, {Vignali}, {Zamojski}, {Aldcroft},
  {Aussel}, {Bardelli}, {Bolzonella}, {Cappi}, {Caputi}, {Contini},
  {Finoguenov}, {Fruscione}, {Garilli}, {Impey}, {Iovino}, {Iwasawa},
  {Kampczyk}, {Kartaltepe}, {Kneib}, {Knobel}, {Kovac}, {Lamareille},
  {Leborgne}, {Le Brun}, {Le Fevre}, {Lilly}, {Maier}, {McCracken}, {Pello},
  {Peng}, {Perez-Montero}, {de Ravel}, {Sanders}, {Scodeggio}, {Scoville},
  {Tanaka}, {Taniguchi}, {Tasca}, {de la Torre}, {Tresse}, {Vergani}, \&
  {Zucca}}]{Brusa:2010lr}
{Brusa}, M., {Civano}, F., {Comastri}, A.,et al.,  2010, \apj, 716, 348

\bibitem[{{Bruzual} \& {Charlot}(2003)}]{Bruzual:2003uq}
{Bruzual}, G. \& {Charlot}, S. 2003, \mnras, 344, 1000

\bibitem[{{Burgett} \& {Kaiser}(2009)}]{Burgett:2009fr}
{Burgett}, W. \& {Kaiser}, N. 2009, in Advanced Maui Optical and Space
  Surveillance Technologies Conference,

\bibitem[{{Calzetti} {et~al.}(2000)}]{Calzetti:2000rt}
{Calzetti}, D., {Armus}, L., {Bohlin}, R.~C., 2000, \apj, 533, 682 

\bibitem[{{Capak} {et~al.}(2007){Capak}, {Aussel}, {Ajiki}, {McCracken},
  {Mobasher}, {Scoville}, {Shopbell}, {Taniguchi}, {Thompson}, {Tribiano},
  {Sasaki}, {Blain}, {Brusa}, {Carilli}, {Comastri}, {Carollo}, {Cassata},
  {Colbert}, {Ellis}, {Elvis}, {Giavalisco}, {Green}, {Guzzo}, {Hasinger},
  {Ilbert}, {Impey}, {Jahnke}, {Kartaltepe}, {Kneib}, {Koda}, {Koekemoer},
  {Komiyama}, {Leauthaud}, {Lefevre}, {Lilly}, {Liu}, {Massey}, {Miyazaki},
  {Murayama}, {Nagao}, {Peacock}, {Pickles}, {Porciani}, {Renzini}, {Rhodes},
  {Rich}, {Salvato}, {Sanders}, {Scarlata}, {Schiminovich}, {Schinnerer},
  {Scodeggio}, {Sheth}, {Shioya}, {Tasca}, {Taylor}, {Yan}, \&
  {Zamorani}}]{Capak:2007db}
{Capak}, P., {Aussel}, H., {Ajiki}, M., et al., 2007, \apjs, 172, 99

\bibitem[{{Cappelluti} {et~al.}(2009){Cappelluti}, {Brusa}, {Hasinger},
  {Comastri}, {Zamorani}, {Finoguenov}, {Gilli}, {Puccetti}, {Miyaji},
  {Salvato}, {Vignali}, {Aldcroft}, {B{\"o}hringer}, {Brunner}, {Civano},
  {Elvis}, {Fiore}, {Fruscione}, {Griffiths}, {Guzzo}, {Iovino}, {Koekemoer},
  {Mainieri}, {Scoville}, {Shopbell}, {Silverman}, \&
  {Urry}}]{Cappelluti:2009fj}
{Cappelluti}, N., {Brusa}, M., {Hasinger}, G., et al., 2009, \aap, 497, 63


\bibitem[{{Cappelluti} {et~al.}(2011){Cappelluti}, {Predehl}, {B{\"o}hringer},
  {Brunner}, {Brusa}, {Burwitz}, {Churazov}, {Dennerl}, {Finoguenov},
  {Freyberg}, {Friedrich}, {Hasinger}, {Kenziorra}, {Kreykenbohm}, {Lamer},
  {Meidinger}, {M{\"u}hlegger}, {Pavlinsky}, {Robrade}, {Santangelo},
  {Schmitt}, {Schwope}, {Steinmitz}, {Str{\"u}der}, {Sunyaev}, \&
  {Tenzer}}]{Cappelluti:2011qy}
{Cappelluti}, N., {Predehl}, P., {B{\"o}hringer}, H., et al.,  2011,
  Memorie della Societa Astronomica Italiana Supplementi, 17, 159 


\bibitem[{{Cardamone} {et~al.}(2010){Cardamone}, {van Dokkum}, {Urry},
  {Taniguchi}, {Gawiser}, {Brammer}, {Taylor}, {Damen}, {Treister}, {Cobb},
  {Bond}, {Schawinski}, {Lira}, {Murayama}, {Saito}, \&
  {Sumikawa}}]{Cardamone:2010lr}
{Cardamone}, C.~N., {van Dokkum}, P.~G., {Urry}, C.~M.,  et al., 2010, \apjs, 189, 270


\bibitem[{{Civano} {et~al.}(2011){Civano}, {Brusa}, {Comastri}, {Elvis},
  {Salvato}, {Zamorani}, {Capak}, {Fiore}, {Gilli}, {Hao}, {Ikeda}, {Kakazu},
  {Kartaltepe}, {Masters}, {Miyaji}, {Mignoli}, {Puccetti}, {Shankar},
  {Silverman}, {Vignali}, \& {Zezas}}]{Civano:2011qy}
{Civano}, F., {Brusa}, M., {Comastri}, A., et al., 2011, ArXiv e-prints


\bibitem[{{Cuillandre} \& {Bertin}(2006)}]{Cuillandre:2006rt}
{Cuillandre}, J.-C. \& {Bertin}, E. 2006, in SF2A-2006: Semaine de
  l'Astrophysique Francaise, ed. {D.~Barret, F.~Casoli, G.~Lagache,
  A.~Lecavelier, \& L.~Pagani }, 265--+

\bibitem[{{Davis} {et~al.}(2007){Davis}, {Guhathakurta}, {Konidaris}, {Newman},
  {Ashby}, {Biggs}, {Barmby}, {Bundy}, {Chapman}, {Coil}, {Conselice},
  {Cooper}, {Croton}, {Eisenhardt}, {Ellis}, {Faber}, {Fang}, {Fazio},
  {Georgakakis}, {Gerke}, {Goss}, {Gwyn}, {Harker}, {Hopkins}, {Huang},
  {Ivison}, {Kassin}, {Kirby}, {Koekemoer}, {Koo}, {Laird}, {Le Floc'h}, {Lin},
  {Lotz}, {Marshall}, {Martin}, {Metevier}, {Moustakas}, {Nandra}, {Noeske},
  {Papovich}, {Phillips}, {Rich}, {Rieke}, {Rigopoulou}, {Salim},
  {Schiminovich}, {Simard}, {Smail}, {Small}, {Weiner}, {Willmer}, {Willner},
  {Wilson}, {Wright}, \& {Yan}}]{Davis:2007uq}
{Davis}, M., {Guhathakurta}, P., {Konidaris}, N.~P., et al.,  2007, \apjl, 660, L1

\bibitem[{{DePoy} {et~al.}(2008){DePoy}, {Abbott}, {Annis}, {Antonik},
  {Barcel{\'o}}, {Bernstein}, {Bigelow}, {Brooks}, {Buckley-Geer}, {Campa},
  {Cardiel}, {Castander}, {Castilla}, {Cease}, {Chappa}, {Dede}, {Derylo},
  {Diehl}, {Doel}, {DeVicente}, {Estrada}, {Finley}, {Flaugher}, {Gaztanaga},
  {Gerdes}, {Gladders}, {Guarino}, {Gutierrez}, {Hamilton}, {Haney}, {Holland},
  {Honscheid}, {Huffman}, {Karliner}, {Kau}, {Kent}, {Kozlovsky}, {Kubik},
  {Kuehn}, {Kuhlmann}, {Kuk}, {Leger}, {Lin}, {Martinez}, {Martinez},
  {Merritt}, {Mohr}, {Moore}, {Moore}, {Nord}, {Ogando}, {Olsen}, {Onal},
  {Peoples}, {Qian}, {Roe}, {Sanchez}, {Scarpine}, {Schmidt}, {Schmitt},
  {Schubnell}, {Schultz}, {Selen}, {Shaw}, {Simaitis}, {Slaughter}, {Smith},
  {Spinka}, {Stefanik}, {Stuermer}, {Talaga}, {Tarle}, {Thaler}, {Tucker},
  {Walker}, {Worswick}, \& {Zhao}}]{DePoy:2008fk}
{DePoy}, D.~L., {Abbott}, T., {Annis}, J.,  et al.,  2008, in Presented
  at the Society of Photo-Optical Instrumentation Engineers (SPIE) Conference,
  Vol. 7014, Society of Photo-Optical Instrumentation Engineers (SPIE)
  Conference Series


\bibitem[{{Duval} {et~al.}(2004){Duval}, {Irace}, {Mainzer}, \&
  {Wright}}]{Duval:2004pd}
{Duval}, V.~G., {Irace}, W.~R., {Mainzer}, A.~K., \& {Wright}, E.~L., 2004, in
  Presented at the Society of Photo-Optical Instrumentation Engineers (SPIE)
  Conference, Vol. 5487, Society of Photo-Optical Instrumentation Engineers
  (SPIE) Conference Series, ed. {J.~C.~Mather}, 101--111

\bibitem[{{Elvis} {et~al.}(2009){Elvis}, {Civano}, {Vignali}, {Puccetti},
  {Fiore}, {Cappelluti}, {Aldcroft}, {Fruscione}, {Zamorani}, {Comastri},
  {Brusa}, {Gilli}, {Miyaji}, {Damiani}, {Koekemoer}, {Finoguenov}, {Brunner},
  {Urry}, {Silverman}, {Mainieri}, {Hasinger}, {Griffiths}, {Carollo}, {Hao},
  {Guzzo}, {Blain}, {Calzetti}, {Carilli}, {Capak}, {Ettori}, {Fabbiano},
  {Impey}, {Lilly}, {Mobasher}, {Rich}, {Salvato}, {Sanders}, {Schinnerer},
  {Scoville}, {Shopbell}, {Taylor}, {Taniguchi}, \& {Volonteri}}]{Elvis:2009kx}
{Elvis}, M., {Civano}, F., {Vignali}, C., et al.  2009, \apjs, 184, 158


\bibitem[{{Fan} {et~al.}(2001){Fan}, {Narayanan}, {Lupton}, {Strauss}, {Knapp},
  {Becker}, {White}, {Pentericci}, {Leggett}, {Haiman}, {Gunn}, {Ivezi{\'c}},
  {Schneider}, {Anderson}, {Brinkmann}, {Bahcall}, {Connolly}, {Csabai}, {Doi},
  {Fukugita}, {Geballe}, {Grebel}, {Harbeck}, {Hennessy}, {Lamb}, {Miknaitis},
  {Munn}, {Nichol}, {Okamura}, {Pier}, {Prada}, {Richards}, {Szalay}, \&
  {York}}]{Fan:2001ys}
{Fan}, X., {Narayanan}, V.~K., {Lupton}, R.~H., et al.,  2001, \aj, 122, 2833 


\bibitem[{{Fan} {et~al.}(2006){Fan}, {Strauss}, {Becker}, {White}, {Gunn},
  {Knapp}, {Richards}, {Schneider}, {Brinkmann}, \& {Fukugita}}]{Fan:2006fk}
{Fan}, X., {Strauss}, M.~A., {Becker}, R.~H.,et al., 2006, \aj, 132, 117


\bibitem[{{Fontanot} {et~al.}(2007){Fontanot}, {Cristiani}, {Monaco}, {Nonino},
  {Vanzella}, {Brandt}, {Grazian}, \& {Mao}}]{Fontanot:2007lr}
{Fontanot}, F., {Cristiani}, S., {Monaco}, P., et al.,  2007, \aap, 461, 39

\bibitem[{{Giavalisco} {et~al.}(2004){Giavalisco}, {Ferguson}, {Koekemoer},
  {Dickinson}, {Alexander}, {Bauer}, {Bergeron}, {Biagetti}, {Brandt},
  {Casertano}, {Cesarsky}, {Chatzichristou}, {Conselice}, {Cristiani}, {Da
  Costa}, {Dahlen}, {de Mello}, {Eisenhardt}, {Erben}, {Fall}, {Fassnacht},
  {Fosbury}, {Fruchter}, {Gardner}, {Grogin}, {Hook}, {Hornschemeier}, {Idzi},
  {Jogee}, {Kretchmer}, {Laidler}, {Lee}, {Livio}, {Lucas}, {Madau},
  {Mobasher}, {Moustakas}, {Nonino}, {Padovani}, {Papovich}, {Park},
  {Ravindranath}, {Renzini}, {Richardson}, {Riess}, {Rosati}, {Schirmer},
  {Schreier}, {Somerville}, {Spinrad}, {Stern}, {Stiavelli}, {Strolger},
  {Urry}, {Vandame}, {Williams}, \& {Wolf}}]{Giavalisco:2004fj}
{Giavalisco}, M., {Ferguson}, H.~C., {Koekemoer}, A.~M., et al.,  2004, \apjl, 600, L93

\bibitem[{{Grogin} {et~al.}(2011){Grogin}, {Kocevski}, {Faber}, {Ferguson},
  {Koekemoer}, {Riess}, {Acquaviva}, {Alexander}, {Almaini}, {Ashby}, {Barden},
  {Bell}, {Bournaud}, {Brown}, {Caputi}, {Casertano}, {Cassata}, {Challis},
  {Chary}, {Cheung}, {Cirasuolo}, {Conselice}, {Roshan Cooray}, {Croton},
  {Daddi}, {Dahlen}, {Dav{\'e}}, {de Mello}, {Dekel}, {Dickinson}, {Dolch},
  {Donley}, {Dunlop}, {Dutton}, {Elbaz}, {Fazio}, {Filippenko}, {Finkelstein},
  {Fontana}, {Gardner}, {Garnavich}, {Gawiser}, {Giavalisco}, {Grazian}, {Guo},
  {Hathi}, {H{\"a}ussler}, {Hopkins}, {Huang}, {Huang}, {Jha}, {Kartaltepe},
  {Kirshner}, {Koo}, {Lai}, {Lee}, {Li}, {Lotz}, {Lucas}, {Madau}, {McCarthy},
  {McGrath}, {McIntosh}, {McLure}, {Mobasher}, {Moustakas}, {Mozena}, {Nandra},
  {Newman}, {Niemi}, {Noeske}, {Papovich}, {Pentericci}, {Pope}, {Primack},
  {Rajan}, {Ravindranath}, {Reddy}, {Renzini}, {Rix}, {Robaina}, {Rodney},
  {Rosario}, {Rosati}, {Salimbeni}, {Scarlata}, {Siana}, {Simard}, {Smidt},
  {Somerville}, {Spinrad}, {Straughn}, {Strolger}, {Telford}, {Teplitz},
  {Trump}, {van der Wel}, {Villforth}, {Wechsler}, {Weiner}, {Wiklind}, {Wild},
  {Wilson}, {Wuyts}, {Yan}, \& {Yun}}]{Grogin:2011fr}
{Grogin}, N.~A., {Kocevski}, D.~D., {Faber}, S.~M., et al., 2011, ArXiv e-prints

\bibitem[{{Hasinger} {et~al.}(2007){Hasinger}, {Cappelluti}, {Brunner},
  {Brusa}, {Comastri}, {Elvis}, {Finoguenov}, {Fiore}, {Franceschini}, {Gilli},
  {Griffiths}, {Lehmann}, {Mainieri}, {Matt}, {Matute}, {Miyaji}, {Molendi},
  {Paltani}, {Sanders}, {Scoville}, {Tresse}, {Urry}, {Vettolani}, \&
  {Zamorani}}]{Hasinger:2007dn}
{Hasinger}, G., {Cappelluti}, N., {Brunner}, H., et al.,  2007, \apjs, 172, 29

\bibitem[{{Hildebrandt} {et~al.}(2008){Hildebrandt}, {Wolf}, \&
  {Ben{\'{\i}}tez}}]{Hildebrandt:2008uq}
{Hildebrandt}, H., {Wolf}, C., \& {Ben{\'{\i}}tez}, N. 2008, \aap, 480, 703

\bibitem[{{Hoaglin} {et~al.}(1983){Hoaglin}, {Mosteller}, \&
  {Tukey}}]{Hoaglin:1983bh}
{Hoaglin}, D.~C., {Mosteller}, F., \& {Tukey}, J.~W. 1983, {Understanding
  robust and exploratory data anlysis} (Wiley Series in Probability and
  Mathematical Statistics, New York: Wiley, 1983, edited by Hoaglin, David C.;
  Mosteller, Frederick; Tukey, John W.)

\bibitem[{{Hopkins} {et~al.}(2008){Hopkins}, {Hernquist}, {Cox}, \& {Kere{\v
  s}}}]{Hopkins:2008qy}
{Hopkins}, P.~F., {Hernquist}, L., {Cox}, T.~J., \& {Kere{\v s}}, D. 2008,
  \apjs, 175, 356

\bibitem[{{Ilbert} {et al.}(2006)}]{Ilbert:2006vl}
{Ilbert}, O., {Arnouts}, S., {McCracken}, H.~J., et al. 2006, \aap, 457,  841

\bibitem[{{Ilbert} {et~al.}(2009){Ilbert}, {Capak}, {Salvato}, {Aussel},
  {McCracken}, {Sanders}, {Scoville}, {Kartaltepe}, {Arnouts}, {Le Floc'h},
  {Mobasher}, {Taniguchi}, {Lamareille}, {Leauthaud}, {Sasaki}, {Thompson},
  {Zamojski}, {Zamorani}, {Bardelli}, {Bolzonella}, {Bongiorno}, {Brusa},
  {Caputi}, {Carollo}, {Contini}, {Cook}, {Coppa}, {Cucciati}, {de la Torre},
  {de Ravel}, {Franzetti}, {Garilli}, {Hasinger}, {Iovino}, {Kampczyk},
  {Kneib}, {Knobel}, {Kovac}, {Le Borgne}, {Le Brun}, {F{\`e}vre}, {Lilly},
  {Looper}, {Maier}, {Mainieri}, {Mellier}, {Mignoli}, {Murayama}, {Pell{\`o}},
  {Peng}, {P{\'e}rez-Montero}, {Renzini}, {Ricciardelli}, {Schiminovich},
  {Scodeggio}, {Shioya}, {Silverman}, {Surace}, {Tanaka}, {Tasca}, {Tresse},
  {Vergani}, \& {Zucca}}]{Ilbert:2009hl}
{Ilbert}, O., {Capak}, P., {Salvato}, M., et al., 2009, \apj, 690, 1236


\bibitem[{{Ilbert} {et~al.}(2010){Ilbert}, {Salvato}, {Le Floc'h}, {Aussel},
  {Capak}, {McCracken}, {Mobasher}, {Kartaltepe}, {Scoville}, {Sanders},
  {Arnouts}, {Bundy}, {Cassata}, {Kneib}, {Koekemoer}, {Le F{\`e}vre}, {Lilly},
  {Surace}, {Taniguchi}, {Tasca}, {Thompson}, {Tresse}, {Zamojski}, {Zamorani},
  \& {Zucca}}]{Ilbert:2010qy}
{Ilbert}, O., {Salvato}, M., {Le Floc'h}, E., et al., 2010, \apj, 709, 644


\bibitem[{{Ivezic} {et~al.}(2006){Ivezic}, {Tyson}, {Strauss}, {Kahn},
  {Stubbs}, {Pinto}, {Cook}, \& {LSST Collaboration}}]{Ivezic:2006mz}
{Ivezic}, Z., {Tyson}, A.~J., {Strauss}, M.~A., et al., 2006, in Bulletin of the
  American Astronomical Society, Vol.~38, American Astronomical Society Meeting
  Abstracts, 1017

\bibitem[{{Koekemoer} {et~al.}(2004)}]{Koekemoer:2004kx}
{Koekemoer}, A.~M.,  {Alexander}, D.~M.,  {Bauer}, F.~E., et al., 2004, \apjl, 600, L123 

\bibitem[{{Koekemoer} {et~al.}(2007)}]{Koekemoer:2007kx}
{Koekemoer}, A.~M., {Aussel}, H.,  {Calzetti}, D., et al., 2007, \apjs, 172, 196 

\bibitem[{{Koekemoer} {et~al.}(2011){Koekemoer}, {Faber}, {Ferguson}, {Grogin},
  {Kocevski}, {Koo}, {Lai}, {Lotz}, {Lucas}, {McGrath}, {Ogaz}, {Rajan},
  {Riess}, {Rodney}, {Strolger}, {Casertano}, {Dahlen}, {Dickinson}, {Dolch},
  {Fontana}, {Giavalisco}, {Grazian}, {Guo}, {Hathi}, {Huang}, {van der Wel},
  {Yan}, {Acquaviva}, {Almaini}, {Ashby}, {Barden}, {Bell}, {Bournaud},
  {Brown}, {Caputi}, {Cassata}, {Challis}, {Chary}, {Cheung}, {Cirasuolo},
  {Conselice}, {Roshan Cooray}, {Croton}, {Daddi}, {Dav{\'e}}, {de Mello}, {de
  Ravel}, {Dekel}, {Donley}, {Dunlop}, {Dutton}, {Elbaz}, {Fazio},
  {Filippenko}, {Finkelstein}, {Frazer}, {Gardner}, {Garnavich}, {Gawiser},
  {Gruetzbauch}, {Hartley}, {H{\"a}ussler}, {Herrington}, {Hopkins}, {Huang},
  {Jha}, {Johnson}, {Kartaltepe}, {Khostovan}, {Kirshner}, {Lani}, {Lee}, {Li},
  {Madau}, {McCarthy}, {McIntosh}, {McLure}, {McPartland}, {Mobasher},
  {Moreira}, {Mortlock}, {Moustakas}, {Mozena}, {Nandra}, {Newman}, {Nielsen},
  {Niemi}, {Noeske}, {Papovich}, {Pentericci}, {Pope}, {Primack},
  {Ravindranath}, {Reddy}, {Renzini}, {Rix}, {Robaina}, {Rosario}, {Rosati},
  {Salimbeni}, {Scarlata}, {Siana}, {Simard}, {Smidt}, {Snyder}, {Somerville},
  {Spinrad}, {Straughn}, {Telford}, {Teplitz}, {Trump}, {Vargas}, {Villforth},
  {Wagner}, {Wandro}, {Wechsler}, {Weiner}, {Wiklind}, {Wild}, {Wilson},
  {Wuyts}, \& {Yun}}]{Koekemoer:2011gf}
{Koekemoer}, A.~M., {Faber}, S.~M., {Ferguson}, H.~C., et al., 2011, ArXiv e-prints

\bibitem[{{Laird} {et~al.}(2009){Laird}, {Nandra}, {Georgakakis}, {Aird},
  {Barmby}, {Conselice}, {Coil}, {Davis}, {Faber}, {Fazio}, {Guhathakurta},
  {Koo}, {Sarajedini}, \& {Willmer}}]{Laird:2009fj}
{Laird}, E.~S., {Nandra}, K., {Georgakakis}, A., et al., 2009, \apjs, 180, 102


\bibitem[{{Leauthaud} {et~al.}(2007){Leauthaud}, {Massey}, {Kneib}, {Rhodes},
  {Johnston}, {Capak}, {Heymans}, {Ellis}, {Koekemoer}, {Le F{\`e}vre},
  {Mellier}, {R{\'e}fr{\'e}gier}, {Robin}, {Scoville}, {Tasca}, {Taylor}, \&
  {Van Waerbeke}}]{Leauthaud:2007fj}
{Leauthaud}, A., {Massey}, R., {Kneib}, J.-P., et al.,  2007, \apjs, 172, 219
{Rhodes}, J., {Johnston}, D.~E.,


\bibitem[{{Lehmer} {et~al.}(2005){Lehmer}, {Brandt}, {Alexander}, {Bauer},
  {Schneider}, {Tozzi}, {Bergeron}, {Garmire}, {Giacconi}, {Gilli}, {Hasinger},
  {Hornschemeier}, {Koekemoer}, {Mainieri}, {Miyaji}, {Nonino}, {Rosati},
  {Silverman}, {Szokoly}, \& {Vignali}}]{Lehmer:2005yq}
{Lehmer}, B.~D., {Brandt}, W.~N., {Alexander}, D.~M., et al.,  2005, \apjs, 161, 2


\bibitem[{{Lilly} {et~al.}(2009){Lilly}, {Le Brun}, {Maier}, {Mainieri},
  {Mignoli}, {Scodeggio}, {Zamorani}, {Carollo}, {Contini}, {Kneib}, {Le
  F{\`e}vre}, {Renzini}, {Bardelli}, {Bolzonella}, {Bongiorno}, {Caputi},
  {Coppa}, {Cucciati}, {de la Torre}, {de Ravel}, {Franzetti}, {Garilli},
  {Iovino}, {Kampczyk}, {Kovac}, {Knobel}, {Lamareille}, {Le Borgne}, {Pello},
  {Peng}, {P{\'e}rez-Montero}, {Ricciardelli}, {Silverman}, {Tanaka}, {Tasca},
  {Tresse}, {Vergani}, {Zucca}, {Ilbert}, {Salvato}, {Oesch}, {Abbas},
  {Bottini}, {Capak}, {Cappi}, {Cassata}, {Cimatti}, {Elvis}, {Fumana},
  {Guzzo}, {Hasinger}, {Koekemoer}, {Leauthaud}, {Maccagni}, {Marinoni},
  {McCracken}, {Memeo}, {Meneux}, {Porciani}, {Pozzetti}, {Sanders},
  {Scaramella}, {Scarlata}, {Scoville}, {Shopbell}, \&
  {Taniguchi}}]{Lilly:2009qy}
{Lilly}, S.~J., {Le Brun}, V., {Maier}, C., et al.,  2009, \apjs, 184, 218


\bibitem[{{Lilly} {et~al.}(2007){Lilly}, {Le F{\`e}vre}, {Renzini}, {Zamorani},
  {Scodeggio}, {Contini}, {Carollo}, {Hasinger}, {Kneib}, {Iovino}, {Le Brun},
  {Maier}, {Mainieri}, {Mignoli}, {Silverman}, {Tasca}, {Bolzonella},
  {Bongiorno}, {Bottini}, {Capak}, {Caputi}, {Cimatti}, {Cucciati}, {Daddi},
  {Feldmann}, {Franzetti}, {Garilli}, {Guzzo}, {Ilbert}, {Kampczyk}, {Kovac},
  {Lamareille}, {Leauthaud}, {Borgne}, {McCracken}, {Marinoni}, {Pello},
  {Ricciardelli}, {Scarlata}, {Vergani}, {Sanders}, {Schinnerer}, {Scoville},
  {Taniguchi}, {Arnouts}, {Aussel}, {Bardelli}, {Brusa}, {Cappi}, {Ciliegi},
  {Finoguenov}, {Foucaud}, {Franceschini}, {Halliday}, {Impey}, {Knobel},
  {Koekemoer}, {Kurk}, {Maccagni}, {Maddox}, {Marano}, {Marconi}, {Meneux},
  {Mobasher}, {Moreau}, {Peacock}, {Porciani}, {Pozzetti}, {Scaramella},
  {Schiminovich}, {Shopbell}, {Smail}, {Thompson}, {Tresse}, {Vettolani},
  {Zanichelli}, \& {Zucca}}]{Lilly:2007qr}
{Lilly}, S.~J., {Le F{\`e}vre}, O., {Renzini}, A., et al.,  2007, \apjs, 172, 70
 

\bibitem[{{Luo} {et~al.}(2010){Luo}, {Brandt}, {Xue}, {Brusa}, {Alexander},
  {Bauer}, {Comastri}, {Koekemoer}, {Lehmer}, {Mainieri}, {Rafferty},
  {Schneider}, {Silverman}, \& {Vignali}}]{Luo:2010qp}
{Luo}, B., {Brandt}, W.~N., {Xue}, et al.,  2010, \apjs, 187, 560


\bibitem[{{Maccacaro} {et~al.}(1988){Maccacaro}, {Gioia}, {Wolter}, {Zamorani},
  \& {Stocke}}]{Maccacaro:1988lr}
{Maccacaro}, T., {Gioia}, I.~M., {Wolter}, A., {Zamorani}, G., \& {Stocke},
  J.~T. 1988, \apj, 326, 680

\bibitem[{{McCracken} {et~al.}(2010){McCracken}, {Capak}, {Salvato}, {Aussel},
  {Thompson}, {Daddi}, {Sanders}, {Kneib}, {Willott}, {Mancini}, {Renzini},
  {Cook}, {Le F{\`e}vre}, {Ilbert}, {Kartaltepe}, {Koekemoer}, {Mellier},
  {Murayama}, {Scoville}, {Shioya}, \& {Tanaguchi}}]{McCracken:2010pj}
{McCracken}, H.~J., {Capak}, P., {Salvato}, M., et al., 2010, \apj, 708, 202


\bibitem[{{Menci} {et~al.}(2008){Menci}, {Fiore}, {Puccetti}, \&
  {Cavaliere}}]{Menci:2008uq}
{Menci}, N., {Fiore}, F., {Puccetti}, S., \& {Cavaliere}, A. 2008, \apj, 686,
  219
  
  \bibitem[{{Mohr} {et~al.}(2008){Mohr}, {Adams}, {Barkhouse}, {Beldica},
  {Bertin}, {Cai}, {da Costa}, {Darnell}, {Daues}, {Jarvis}, {Gower}, {Lin},
  {Martelli}, {Neilsen}, {Ngeow}, {Ogando}, {Parga}, {Sheldon}, {Tucker},
  {Kuropatkin}, \& {Stoughton}}]{Mohr:2008qy}
{Mohr}, J.~J., {Adams}, D., {Barkhouse}, W.,  et al.,  2008, in Presented at the Society of
  Photo-Optical Instrumentation Engineers (SPIE) Conference, Vol. 7016, Society
  of Photo-Optical Instrumentation Engineers (SPIE) Conference Series


\bibitem[{{Norris}(2010)}]{Norris:2010qf}
{Norris}, R. 2010, in Bulletin of the American Astronomical Society, Vol.~36,
  American Astronomical Society Meeting Abstracts, 604.05--+

\bibitem[{{Oyaizu} {et~al.}(2008){Oyaizu}, {Lima}, {Cunha}, {Lin}, {Frieman},
  \& {Sheldon}}]{Oyaizu:2008lr}
{Oyaizu}, H., {Lima}, M., {Cunha}, C.~E., {Lin}, H., {Frieman}, J., \&
  {Sheldon}, E.~S. 2008, \apj, 674, 768

\bibitem[{{Polletta} {et~al.}(2007)}]{Polletta:2007cs}
{Polletta}, M., {Tajer}, M., {Maraschi}, L., et al., 2007, \apj, 663, 81 


\bibitem[{{Predehl} {et~al.}(2007){Predehl}, {Andritschke}, {Bornemann},
  {Br{\"a}uninger}, {Briel}, {Brunner}, {Burkert}, {Dennerl}, {Eder},
  {Freyberg}, {Friedrich}, {F{\"u}rmetz}, {Hartmann}, {Hartner}, {Hasinger},
  {Herrmann}, {Holl}, {Huber}, {Kendziorra}, {Kink}, {Meidinger}, {M{\"u}ller},
  {Pavlinsky}, {Pfeffermann}, {Roh{\'e}}, {Santangelo}, {Schmitt}, {Schwope},
  {Steinmetz}, {Str{\"u}der}, {Sunyaev}, {Tiedemann}, {Vongehr}, {Wilms},
  {Erhard}, {Gutruf}, {Jugler}, {Kampf}, {Graue}, {Citterio}, {Valsecci},
  {Vernani}, \& {Zimmerman}}]{Predehl:2007gf}
{Predehl}, P., {Andritschke}, R., {Bornemann}, W., et al.,  2007, in Presented at the Society of
  Photo-Optical Instrumentation Engineers (SPIE) Conference, Vol. 6686, Society
  of Photo-Optical Instrumentation Engineers (SPIE) Conference Serie


\bibitem[{{Prescott} {et~al.}(2006){Prescott}, {Impey}, {Cool}, \&
  {Scoville}}]{Prescott:2006ek}
{Prescott}, M.~K.~M., {Impey}, C.~D., {Cool}, R.~J., \& {Scoville}, N.~Z. 2006,
  \apj, 644, 100

\bibitem[{{Prevot} {et~al.}(1984)}]{Prevot:1984bf}
{Prevot}, M.~L., {Lequeux}, J., {Prevot}, L., {Maurice}, E., {Rocca-Volmerange}, B., 1984, \aap, 132, 389

\bibitem[{{Puccetti} {et~al.}(2009){Puccetti}, {Vignali}, {Cappelluti},
  {Fiore}, {Zamorani}, {Aldcroft}, {Elvis}, {Gilli}, {Miyaji}, {Brunner},
  {Brusa}, {Civano}, {Comastri}, {Damiani}, {Fruscione}, {Finoguenov},
  {Koekemoer}, \& {Mainieri}}]{Puccetti:2009rt}
{Puccetti}, S., {Vignali}, C., {Cappelluti}, N., et al.,  2009, \apjs, 185, 586



\bibitem[{{Rovilos} {et~al.}(2009){Rovilos}, {Burwitz}, {Szokoly}, {Hasinger},
  {Egami}, {Bouch{\'e}}, {Berta}, {Salvato}, {Lutz}, \&
  {Genzel}}]{Rovilos:2009fj}
{Rovilos}, E., {Burwitz}, V., {Szokoly}, G., et al., 2009, \aap, 507, 195


\bibitem[{{Rovilos} {et~al.}(2011){Rovilos}, {Fotopoulou}, {Salvato},
  {Burwitz}, {Egami}, {Hasinger}, \& {Szokoly}}]{Rovilos:2011yq}
{Rovilos}, E., {Fotopoulou}, S., {Salvato}, M., et al., 2011, ArXiv e-prints

\bibitem[{{Salvato} {et~al.}(2009){Salvato}, {Hasinger}, {Ilbert}, {Zamorani},
  {Brusa}, {Scoville}, {Rau}, {Capak}, {Arnouts}, {Aussel}, {Bolzonella},
  {Buongiorno}, {Cappelluti}, {Caputi}, {Civano}, {Cook}, {Elvis}, {Gilli},
  {Jahnke}, {Kartaltepe}, {Impey}, {Lamareille}, {Le Floc'h}, {Lilly},
  {Mainieri}, {McCarthy}, {McCracken}, {Mignoli}, {Mobasher}, {Murayama},
  {Sasaki}, {Sanders}, {Schiminovich}, {Shioya}, {Shopbell}, {Silverman},
  {Smol{\v c}i{\'c}}, {Surace}, {Taniguchi}, {Thompson}, {Trump}, {Urry}, \&
  {Zamojski}}]{Salvato:2009zw}
{Salvato}, M., {Hasinger}, G., {Ilbert}, O., et al., 2009, \apj, 690, 1250


\bibitem[{{Sanders} {et~al.}(2007){Sanders}, {Salvato}, {Aussel}, {Ilbert},
  {Scoville}, {Surace}, {Frayer}, {Sheth}, {Helou}, {Brooke}, {Bhattacharya},
  {Yan}, {Kartaltepe}, {Barnes}, {Blain}, {Calzetti}, {Capak}, {Carilli},
  {Carollo}, {Comastri}, {Daddi}, {Ellis}, {Elvis}, {Fall}, {Franceschini},
  {Giavalisco}, {Hasinger}, {Impey}, {Koekemoer}, {Le F{\`e}vre}, {Lilly},
  {Liu}, {McCracken}, {Mobasher}, {Renzini}, {Rich}, {Schinnerer}, {Shopbell},
  {Taniguchi}, {Thompson}, {Urry}, \& {Williams}}]{Sanders:2007qd}
{Sanders}, D.~B., {Salvato}, M., {Aussel}, H., et al., 2007, \apjs, 172, 86


\bibitem[{{Scoville} {et~al.}(2007){Scoville}, {Abraham}, {Aussel}, {Barnes},
  {Benson}, {Blain}, {Calzetti}, {Comastri}, {Capak}, {Carilli}, {Carlstrom},
  {Carollo}, {Colbert}, {Daddi}, {Ellis}, {Elvis}, {Ewald}, {Fall},
  {Franceschini}, {Giavalisco}, {Green}, {Griffiths}, {Guzzo}, {Hasinger},
  {Impey}, {Kneib}, {Koda}, {Koekemoer}, {Lefevre}, {Lilly}, {Liu},
  {McCracken}, {Massey}, {Mellier}, {Miyazaki}, {Mobasher}, {Mould}, {Norman},
  {Refregier}, {Renzini}, {Rhodes}, {Rich}, {Sanders}, {Schiminovich},
  {Schinnerer}, {Scodeggio}, {Sheth}, {Shopbell}, {Taniguchi}, {Tyson}, {Urry},
  {Van Waerbeke}, {Vettolani}, {White}, \& {Yan}}]{Scoville:2007rw}
{Scoville}, N., {Abraham}, R.~G., {Aussel}, H., et al., 2007, \apjs, 172, 38


\bibitem[{{Taylor}(2005)}]{Taylor:2005rt}
{Taylor}, M.~B. 2005, in Astronomical Society of the Pacific Conference Series,
  Vol. 347, Astronomical Data Analysis Software and Systems XIV, ed.
  {P.~Shopbell, M.~Britton, \& R.~Ebert}, 29--+

\bibitem[{{Tisserand} {et~al.}(2008){Tisserand}, {Keller}, {Schmidt}, \&
  {Bessell}}]{Tisserand:2008fj}
{Tisserand}, P., {Keller}, S., {Schmidt}, B., \& {Bessell}, M. 2008, {SkyMapper
  and the Southern Sky Survey}, ed. {Koribalski, B.~S.~\& Jerjen, H.}, 337
  
\bibitem[{{Trump} {et~al.}(2007){Trump}, {Impey}, {McCarthy}, {Elvis},
  {Huchra}, {Brusa}, {Hasinger}, {Schinnerer}, {Capak}, {Lilly}, \&
  {Scoville}}]{Trump:2007fv}
{Trump}, J.~R., {Impey}, C.~D., {McCarthy}, P.~J., et al.,  2007, \apjs, 172, 383

\bibitem[{{Veron-Cetty} \& {Veron}(1998)}]{Veron-Cetty:1998lr}
{Veron-Cetty}, M.-P. \& {Veron}, P. 1998, {A Catalogue of quasars and active
  nuclei}, ed. {Veron-Cetty, M.-P.~\& Veron, P.}


\bibitem[{{Williams} {et~al.}(1996){Williams}, {Blacker}, {Dickinson}, {Dixon},
  {Ferguson}, {Fruchter}, {Giavalisco}, {Gilliland}, {Heyer}, {Katsanis},
  {Levay}, {Lucas}, {McElroy}, {Petro}, {Postman}, {Adorf}, \&
  {Hook}}]{Williams:1996lr}
{Williams}, R.~E., {Blacker}, B., {Dickinson}, M., et al.,1996, \aj, 112, 1335


\bibitem[{{Willott} {et~al.}(2003){Willott}, {McLure}, \&
  {Jarvis}}]{Willott:2003qy}
{Willott}, C.~J., {McLure}, R.~J., \& {Jarvis}, M.~J. 2003, \apjl, 587, L15

\bibitem[{{Wolf} {et~al.}(2003){Wolf}, {Meisenheimer}, {Rix}, {Borch}, {Dye},
  \& {Kleinheinrich}}]{Wolf:2003fk}
{Wolf}, C., {Meisenheimer}, K., {Rix}, H., {Borch}, A., {Dye}, S., \&
  {Kleinheinrich}, M. 2003, \aap, 401, 73

\bibitem[{{Wolf} {et~al.}(2001){Wolf}, {Meisenheimer}, {R{\"o}ser}, {Beckwith},
  {Chaffee}, {Fried}, {Hippelein}, {Huang}, {K{\"u}mmel}, {von Kuhlmann},
  {Maier}, {Phleps}, {Rix}, {Thommes}, \& {Thompson}}]{Wolf:2001ul}
{Wolf}, C., {Meisenheimer}, K., {R{\"o}ser}, H.-J., et al., 2001, \aap, 365, 681


\bibitem[{{Wright} {et~al.}(2010){Wright}, {Drake}, \&
  {Civano}}]{Wright:2010qy}
{Wright}, N.~J., {Drake}, J.~J., \& {Civano}, F. 2010, \apj, 725, 480

\bibitem[{{Zamojski} {et~al.}(2007){Zamojski}, {Schiminovich}, {Rich},
  {Mobasher}, {Koekemoer}, {Capak}, {Taniguchi}, {Sasaki}, {McCracken},
  {Mellier}, {Bertin}, {Aussel}, {Sanders}, {Le F{\`e}vre}, {Ilbert},
  {Salvato}, {Thompson}, {Kartaltepe}, {Scoville}, {Barlow}, {Forster},
  {Friedman}, {Martin}, {Morrissey}, {Neff}, {Seibert}, {Small}, {Wyder},
  {Bianchi}, {Donas}, {Heckman}, {Lee}, {Madore}, {Milliard}, {Szalay},
  {Welsh}, \& {Yi}}]{Zamojski:2007lq}
{Zamojski}, M.~A., {Schiminovich}, D., {Rich}, R.~M., et al., 2007, \apjs, 172, 468


\end{thebibliography}
{}

\begin{table*}[HT!]
\caption{Assessing quality of photo-z for the  {\it EXTNV} sub-samples using different libraries}
 \begin{center}
 \begin{tabular}{l|ll|ll|ll|ll}
                                 
\hline
                                        &  \multicolumn{6}{c}{C-COSMOS {\it EXTNV}} &    \multicolumn{2}{|c}{XMM-COSMOS {\it EXTNV}}\\
&  \multicolumn{2}{c}{i$<22.5$} &    \multicolumn{2}{c}{i$>22.5$}&      \multicolumn{2}{c|}{all}&      \multicolumn{2}{c}{i$<$22.5}  \\
 
Library & $\eta$ (\%) &$\sigma_{NMAD}$  &  $\eta$ (\%) &$\sigma_{NMAD}$   &  $\eta$ (\%) &$\sigma_{NMAD}$ &  $\eta$ (\%) &$\sigma_{NMAD}$\\
\hline
\hline
I09\tablenotemark{1}               & 4.2 & 0.015 & 9.0   & 0.041 & 5.7 & 0.017  & 7.7& 0.017\\
S09\tablenotemark{2}              &2.7 & 0.020 & 18.4 & 0.083 & 6.0& 0.028 & 4.4 & 0.022\\ 
{\bf Combined\tablenotemark{3}}    & {\bf 2.4} & {\bf 0.014} & {\bf 10.3}   & {\bf 0.041} & {\bf 4.1} & {\bf 0.017} & {\bf4.7} & {\bf0.016}\\
\end{tabular}
\end{center}
\tablenotetext{1}{{Library} from I09; only normal galaxy templates.}
\tablenotetext{2}{{Library} from S09; mostly AGN-dominated templates.}
\tablenotetext{3}{Final result obtained using I09 or S09 library, depending on the X-ray flux of the sources.}

\label{tab:extnv}
\end{table*}

\begin{table}[htdp]

\caption{Results for the C-COSMOS {\it QSOV} sample using S09 and I09 libraries  }
 \begin{center}
\begin{tabular}{l|ll|ll|ll}    
\small
Library     &   \multicolumn{2}{c|}{{\it QSOV}, $i^*<=22.5$}  &  \multicolumn{2}{c|}{{\it QSOV}, $i^*>22.5$}   & \multicolumn{2}{c}{{\it EXTNV}, all}    \\     
                 & $\eta$(\%) & $\sigma_{NMAD}$                      &  $\eta$(\%) & $\sigma_{NMAD}$                    &    $\eta$(\%) & $\sigma_{NMAD}$  \\
\hline
                
&&&&&& \\
I09       & 45.6 & 0.165   & 23.5 &0.074 & 38.2 & 0.135\\

 S09   &5.1 & 0.011    & 14.3 &0.022 & 8.2  & 0.013 \\             

\end{tabular}
\end{center}
\label{tab:qsov}
\end{table}

\begin{table*}[htdp]

\caption{XMM-COSMOS recomputed}
 \begin{center}
\begin{tabular}{l|ccc|ccc||ccc}
\small       &        \multicolumn{6}{c|}{ i$^*<$22.5}           &      \multicolumn{3}{|c}{total}  \\     
               &   \multicolumn{3}{c}{{\it EXTNV}} &   \multicolumn{3}{|c|}{{\it QSOV}}  & \multicolumn{3}{|c}{{\it QSOV}+ {\it EXTNV}} \\
 Library            & N\tablenotemark{1} & $\eta$  & $\sigma_{NMAD}$ & N\tablenotemark{1}  & $\eta$  &$\sigma_{NMAD}$  & N\tablenotemark{1}  &$\eta$ &$\sigma_{NMAD}$ \\  
\hline
&&&&&&&&&\\
results from S09     & 218 & 2.3\% & 0.019 & 178 & 6.3\% & 0.012 & 442 & 5.3\% &0.017 \\
 &&&&&&\\
&&&&&&\\
same procedure as S09             & 270 & 4.4\% & 0.022 & 236 & 7.2\% & 0.013 & 590 & 6.3\% & 0.017 \\
with addition of H-band &&&&&&\\
and more spectroscopy &&&&&&\\
&&&&&&\\
new method     &  270 & {\bf 4.1\%} & {\bf 0.017} & 236 & {\bf 7.2\%} & {\bf 0.013} & 590 & {\bf 6.1\%} &{\bf 0.015}  \\
&&&&&&\\
\end{tabular}
\end{center}
\label{tab:old-new-xmm}
\tablenotetext{1}{number of sources with spectroscopic redshift}
\end{table*}

\begin{table*}[htdp]
\caption{Extracted from C-COSMOS Photometric redshift catalog}

\begin{center}
\begin{tabular}{ccccccccc}

\small XID &  ID(Ilbert)  &  z$_{phot}$    & z$_{phot}$lower  & z$_{phot}$upper    & PDFz & Template &  Morph. & VAR   \\ 
            (1)  &    (2)         &   (3)                &     (4)                    &    (5)                      &  (6)     &     (7)       &    (8)    &     (9)     \\
 \hline
 \hline
 
 1   & 860777   & 1.93 & 1.86 & 2.03   & 86.52  & 3  & -999  &0.48  \\
6    & 1081059  & 1.12 & 1.11 & 1.14   & 100.00 & 1  &  1    &0.83  \\
14  & 1046901  & 2.15 & 2.07 & 2.21   & 95.67  & 22 &  1    &0.30  \\
21  & 1007423  & 1.86 & 1.85 & 1.88   & 100.00 & 30 &  2    &0.28  \\
23  & 997226   & 2.93 & 2.91 & 2.95   & 67.41  & 1  &  1    &0.44  \\
25  & 974083   & 1.99 & 1.95 & 2.04   & 98.81  & 5  & -999  &0.31  \\
26  & 969546   & 0.73 & 0.72 & 0.74   & 100.00 & 28 &  2    &0.57  \\
27  & 974555   & 1.51 & 1.46 & 1.57   & 95.59  & 5  &  1    &0.47  \\
29  & 972975   & 1.08 & 1.05 & 1.10   & 99.76  & 20 &  2    &0.17  \\
31  & 978155   & 2.62 & 2.6  & 2.62   & 100.00 & 26 &  2    &0.06  \\
 \hline
\end{tabular}
\tablecomments{Excerpt from the photo-z catalog available online for C-COSMOS. Column 1:Either {\it Chandra}  from Civano et al. 2011 (in preparation)  or XMM identifier \cite[from][]{Brusa:2010lr}; Column 2 :Optical identifier number as reported in the optical catalog and in Ilbert et al. (2009); Column 3: Photometric redshift; Column 4 and 5: Lower and Upper value of photometric redshift;  Column 6: redshift probability distribution; Column 7:  Best-fit template: from 1 to 30 the templates are from S09, templates from 100+(1...31) are from the I09 library; Column 8: Morphological classification \cite[from][]{Leauthaud:2007fj}  1 or 3 indicates extended sources, while 2 or 4 indicates point-like sources; Column 9: Variability. A revised photo-z catalog for XMM-COSMOS, with the same structure is available at the same address.}
\end{center}
\label{tab:summary_example}
\end{table*}
 
\end{document}